\documentclass[aps,twocolumn,superscriptaddress]{revtex4-1}

\usepackage{graphicx}
\usepackage{amsmath}
\usepackage{amssymb}
\usepackage{amsbsy}
\usepackage{bm}
\usepackage{setspace}
\usepackage{amsfonts}
\usepackage{textcomp}
\usepackage{gensymb}

\begin{document}

\title{Metachronal waves in the flagellar beating of \textit{Volvox}\\ and their hydrodynamic origin}

\author{Douglas R. Brumley}
\affiliation{Department of Applied Mathematics and Theoretical Physics, Centre for Mathematical Sciences, \\
University of Cambridge, Wilberforce Road, Cambridge CB3 0WA, United Kingdom}\affiliation{Department of Civil 
and Environmental Engineering, Massachusetts Institute of Technology, \\ 77 Massachusetts Avenue, Cambridge, 
Massachusetts 02139, USA}
\author{Marco Polin}
\affiliation{Physics Department, University of Warwick, Gibbet Hill Road, Coventry CV4 7AL, United Kingdom}
\author{Timothy J. Pedley}
\affiliation{Department of Applied Mathematics and Theoretical Physics, Centre for Mathematical Sciences, \\
University of Cambridge, Wilberforce Road, Cambridge CB3 0WA, United Kingdom}
\author{Raymond E. Goldstein}
\email{R.E.Goldstein@damtp.cam.ac.uk}
\affiliation{Department of Applied Mathematics and Theoretical Physics, Centre for Mathematical Sciences, \\
University of Cambridge, Wilberforce Road, Cambridge CB3 0WA, United Kingdom}

\date{\today}

\begin{abstract}
Groups of eukaryotic cilia and flagella are capable of coordinating their beating over large scales, routinely 
exhibiting collective dynamics in the form of metachronal waves. The origin of this behaviour -- possibly 
influenced by both mechanical interactions and direct biological regulation -- is poorly understood, in large 
part due to lack of quantitative experimental studies.
Here we characterise in detail flagellar coordination on the surface of the multicellular alga {\it Volvox carteri}, 
an emerging model organism for flagellar dynamics. Our studies reveal for the first time that the average 
metachronal coordination observed is punctuated by periodic phase defects during which synchrony is partial 
and limited to specific groups of cells.
A minimal model of hydrodynamically coupled oscillators can reproduce semi-quantitatively the characteristics 
of the average metachronal dynamics, and the emergence of defects. We systematically study the model's 
behaviour by assessing the effect of changing intrinsic rotor characteristics, including oscillator 
stiffness and the nature of their internal driving force, as well as their geometric properties and 
spatial arrangement.
Our results suggest that metachronal coordination follows from deformations in the oscillators' limit 
cycles induced by hydrodynamic stresses, and that defects result from sufficiently steep local biases 
in the oscillators' intrinsic frequencies. Additionally, we find that random variations in the intrinsic 
rotor frequencies increase the robustness of the average properties of the emergent metachronal waves. \\

\noindent \textbf{Keywords:} eukaryotic flagella, metachronal waves, microhydrodynamics, synchronization, 
phase defects, colloidal oscillators.
\end{abstract}

\maketitle

\section*{Introduction}

The  eukaryotic flagellum is one of the most highly conserved structures in biology, providing locomotion and 
fluid transport through the execution of a periodic motion. From mucociliary clearance in the human respiratory 
tract \cite{Button:2012}, to cerebrospinal flows \cite{Guirao2010}, and the establishment of left-right asymmetry 
in mammalian embryos \cite{Nonaka2002}, the flows generated by this periodic beating are tightly coupled 
to microscale transport in fundamental biological processes.
Physically separated pairs of flagella can synchronize their beating purely through hydrodynamic 
interactions \cite{Brumley2014}, but cilia and flagella usually fulfil their tasks through the concerted 
action of large groups, displaying a remarkably 
universal tendency to develop synchronous beating patterns known as metachronal waves (MWs), large scale 
modulations of the beating phase \cite{Knight_Jones:1954}.
Readily observed on the ciliated surface of protists like {\it Opalina} and {\it Paramecium} \cite{Tamm:1970,Tamm:1972} 
and even in synthetic bundles of active microtubules \cite{Sanchez:2011fk}, MWs are still a mystery: their emergence, 
their properties, and their biological role are poorly understood, mostly due to the difficulty of studying 
them {\it in vivo}. 

Early qualitative studies of MWs \cite{Sleigh:1962,Brennen:1977,Machemer:1972ys} suggested a mechanical 
origin for synchronization, 
and motivated theoretical studies of hydrodynamically coupled filaments driven by different internal 
engines \cite{Machin:1963,Gueron:1997,Gueron:1999,Yang:2008,Elgeti:2013}. While these models show a general 
tendency towards metachronism, their complexity prevents a simple understanding of the underlying mechanism.
Minimal models of hydrodynamically coupled self-sustained oscillators, instead, have the potential to offer insights 
into the emergence of MWs, since in many cases they admit analytical solutions which provide a direct link between model
parameters and details of synchrony. 
Experiments with rotating paddles \cite{Qian:2009}, light driven microrotors \cite{DiLeonardo12}, and colloids in 
optical tweezers \cite{Kotar:2010,Cicuta:2012,Lhermerout:2012}, as well as simulations of rotating helices \cite{Reichert:2005qf}, 
or spheres driven along fixed \cite{Lagomarsino:2003zr,Wollin:2011,Vilfan:2006uq,Lenz:2006fk,Uchida:2011kx} or 
flexible \cite{Niedermayer:2008fk,Brumley:2012} trajectories, have shown that under certain conditions
hydrodynamic interactions alone can induce phase-locking of two simple oscillators. 
This locking is mediated either through wall-modified flows \cite{Vilfan:2006uq}, a variable driving 
force \cite{Uchida:2011kx,Uchida:2012fk}, or mechanical elasticity intrinsic in the system \cite{Niedermayer:2008fk,Brumley:2012}. 
Of these, the last two provide the fastest, and possibly equivalently strong \cite{Kotar2013}, drive towards synchronization.
Meanwhile, experiments on the biflagellate alga {\it Chlamydomonas reihnardtii}
\cite{Goldstein:2009vn,Goldstein:2011_emergence,Polin:2009kx}, and on isolated pairs of somatic cells from the 
multicellular alga {\it Volvox carteri} \cite{Brumley2014}, lend  strong support to the idea that synchronization 
among two flagella happens indeed through the elastic response of flagella to shear stress. Even though two flagella can synchronize their motions through direct hydrodynamic interactions \cite{Brumley2014}, this pair synchronization does not necessarily guarantee group synchronization, nor the emergence of MWs \cite{Niedermayer:2008fk}.
Owing to its large size and ease of visualisation, the colonial alga {\it Volvox carteri} is an ideal model 
organism for the 
study of flagella-driven flows \cite{ARFM}. {\it Volvox} comprises thousands of biflagellate somatic cells, embedded within a 
spherical extracellular matrix shell, beating their flagella towards the colony's posterior with only a small deviation 
($\sim 15^{\circ}$) from meridian lines \cite{Kirk:Volvox}.
Here we show that these cells display a complex dynamical behaviour, with colony-wide MWs propagating towards the posterior interrupted by characteristically shaped recurrent phase defects, not previously observed in experiments.
For realistic parameters, we show that a minimal model of hydrodynamically coupled identical rotors predicts
MWs of wavelength similar to those observed in experiments. These MWs emerge in a manner essentially 
independent of boundary conditions, and are robust against realistic perturbations in the rotors' properties. 
Including an intrinsic frequency bias derived from experiments, the model develops periodic defects similar to those we report for {\it Volvox}.
These findings suggest that the collective dynamics of {\it Volvox} flagella result from a competition between a drive towards synchronization, based on the oscillators' elastic compliance, and additional strains imposed by a large scale bias in flagellar properties.

\section*{Metachronal waves in \textit{Volvox carteri}}

{\it Volvox carteri} f. {\it nagariensis} (strain EVE) was grown axenically in Standard {\it Volvox} 
Medium \cite{Kirk:1983uq,Short:2006,Solari:2006} bubbled with sterile air. The cultures were hosted in a 
growth chamber (Binder, Germany) set to a cycle of 16~h light (100~$\mu$Em$^{-2}$s$^{-1}$, Fluora, OSRAM)
at 28\textdegree C and 8 h dark at 26\textdegree C. Individual {\it Volvox} colonies within a $25\times 25\times 5\,$mm 
glass observation chamber filled with fresh medium were captured, oriented and held in place using two glass 
micropipettes with $100\,\mu$m diameter tips housed in pipette holders (World Precision Instruments, USA) 
connected to two manual microinjectors (Sutter Instruments Co., USA). Motorised micromanipulators (Patchstar, 
Scientifica, USA) and custom-made rotation stages allowed the pipettes to be moved in three dimensions and rotated
freely along their centrelines. 
We aligned the axis of each colony along the focal plane of the $40\times$ Plan Fluor lens (NA 0.6) of a Nikon 
TE2000-U inverted microscope, and recorded $30\,$s long movies of the surrounding flow with a high-speed 
camera (Fastcam SA3, Photron, USA) at $500\,$fps  under bright field illumination. 
A long pass interference filter with a 10~nm transition ramp centred at 620~nm (Knight Optical, UK) prevented 
phototactic responses of the {\it Volvox} colonies. 

\begin{figure}[t!]
\begin{center}
\includegraphics[width=\columnwidth]{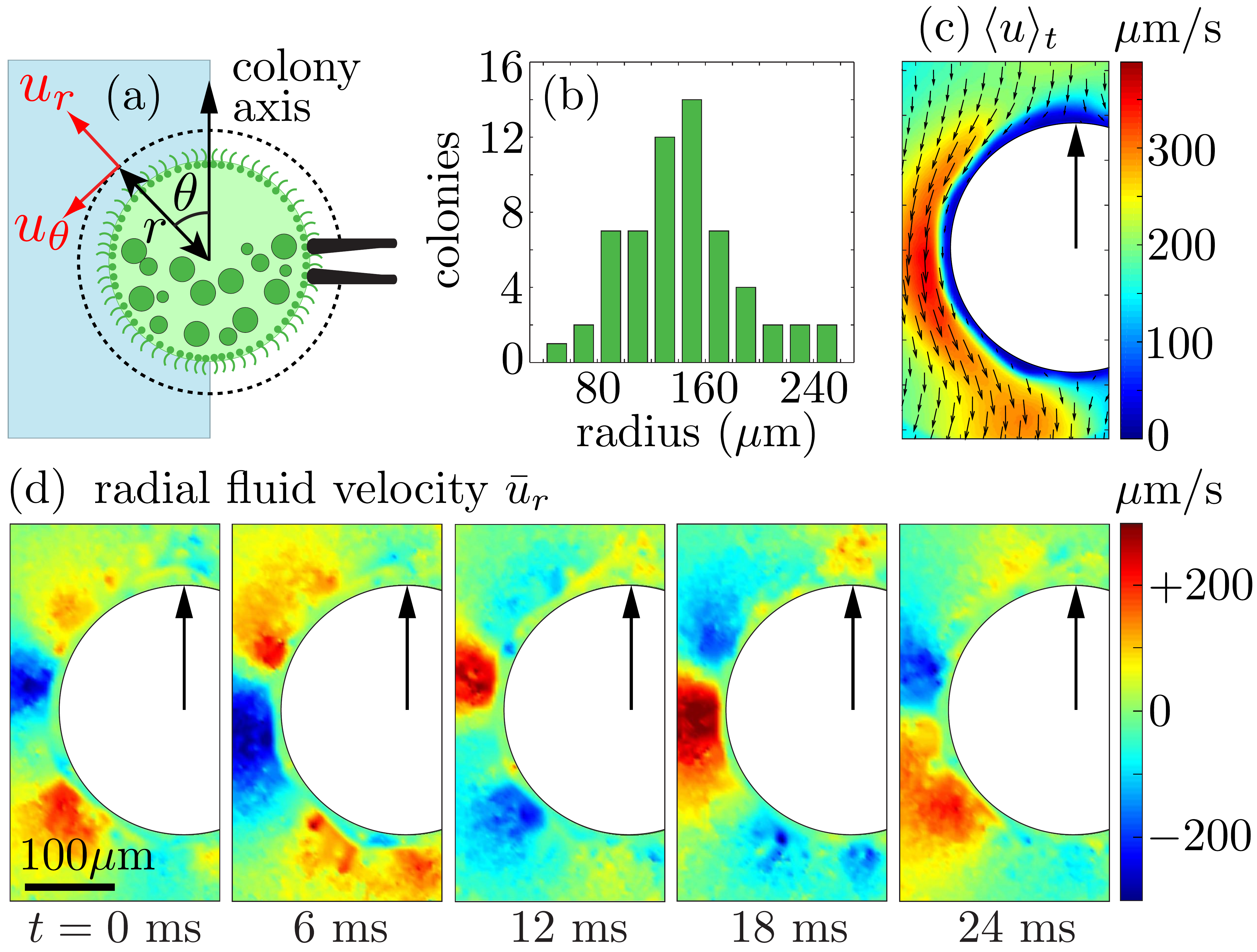}
\caption{{\it Volvox} geometry and flows. (a) A {\it Volvox} colony held by a micropipette showing flagellated somatic cells (small dots) 
and interior daughter colonies (large circles) growing in the posterior half.  The dashed line indicates the distance at which the kymographs 
for the components of the fluid velocity, $u_{r}$ and $u_{\theta}$, are measured. (b) The distribution of radii of {\it Volvox} colonies used 
in experiments (total $n=60$). (c) Magnitude (colour) and direction (vector) of the time-averaged flow field obtained using PIV. (d) Radial 
component of the instantaneous fluid velocity, $\bar{u}_r$, at various times throughout one flagellar beat cycle. The disturbance propagating 
towards the posterior of the colony is clearly visible.}
\label{fig1}
\end{center}
\end{figure}

The projection of the velocity field $\bm{u}$ onto the focal plane was visualised by seeding the fluid with 
$0.5~\mu$m polystyrene microspheres (Invitrogen, USA) at $2 \times 10^{-4}$ volume fraction, and measured 
using an open source Particle Image Velocimetry (PIV) tool for Matlab (MatPIV) (Fig.~\ref{fig1}). It was then decomposed into radial and tangential components 
$\bm{u}(r,\theta,t) = u_{r}(r,\theta,t) \bm{e}_{r} + u_{\theta}(r,\theta,t) \bm{e}_{\theta}$, where the local 
coordinates are defined with respect to the centre of the {\it Volvox} colony as shown in 
Fig.~\ref{fig1}(a).  A total of 60 {\it Volvox} colonies were selected at random 
from the stock supply, and varied in radius $R$ from $48$ to $251\,\mu$m (mean $144 \pm 43\,\mu$m), with a distribution shown in Fig.~\ref{fig1}(b). 

\begin{figure}[t!]
\begin{center}
\includegraphics[width=\columnwidth]{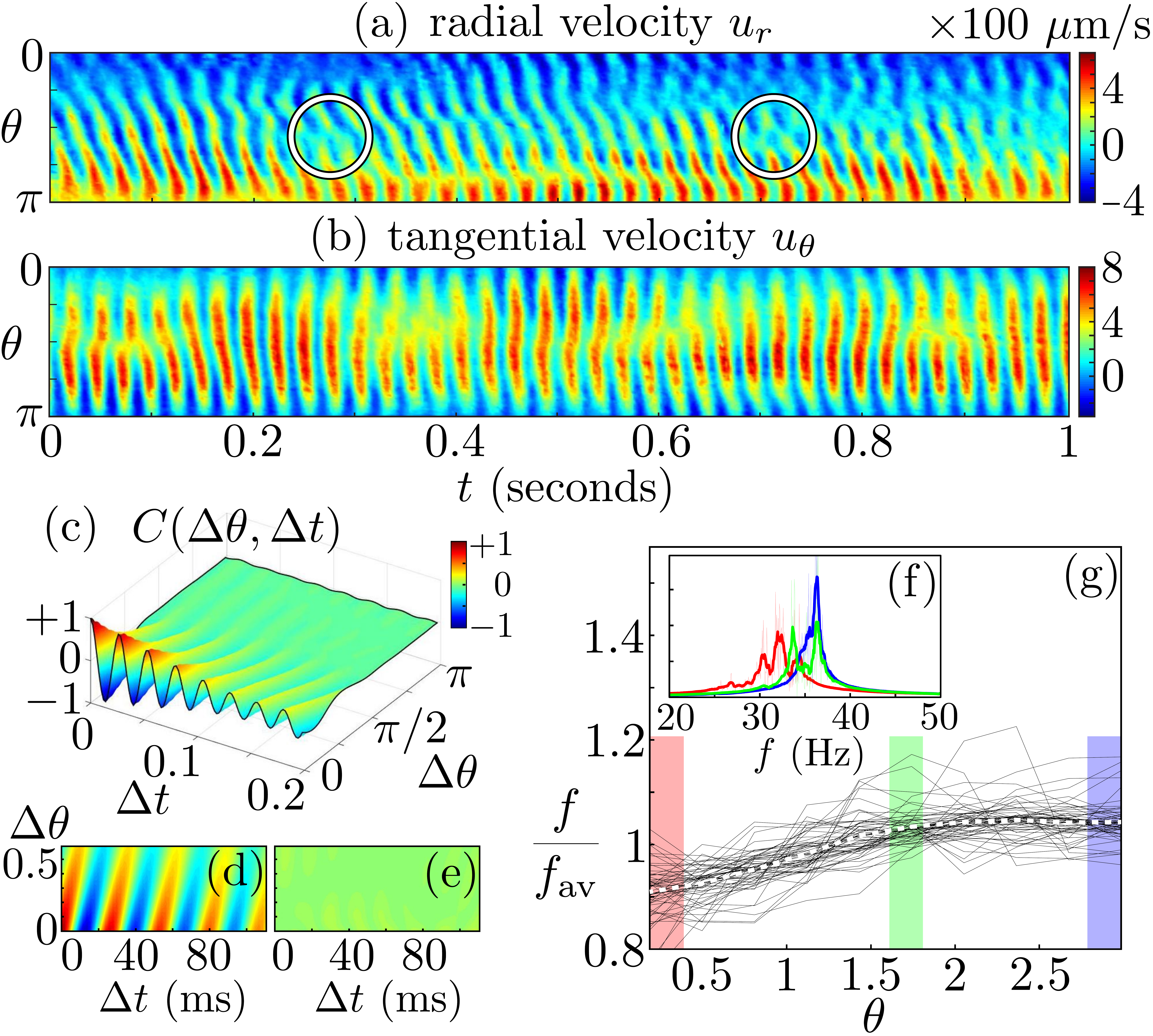}
 \caption{Properties of metachronal waves. (a) Radial $u_{r}(r^*,\theta,t)$ and (b) tangential 
$u_{\theta}(r^*,\theta,t)$ components of the flow field, measured at $r^{*}=1.3\,R$. Phase defects are evidenced by white circles.
(c) Correlation function $C(\Delta \theta, \Delta t)$ for a single representative {\it Volvox} colony, (d) its fitted 
correlation function $C^{\text{fit}}(\Delta \theta, \Delta t) = e^{-\Delta t/\tau} 
e^{-\Delta\theta/L} \cos(k\,\Delta\theta - \omega\,\Delta t)$ and the (e) corresponding error. The scale bar 
for (c-e) is the same. (f) Power spectrum of the autocorrelation function $C(0,\Delta t)$ (for one {\it Volvox} colony), 
calculated for three distinct values of $\theta$. (g) Average beat frequency as a function of polar angle $\theta$ for 
$n=60$ different colonies (black) as well as the ensemble average (white dotted).}
\label{fig2}
\end{center}
\end{figure}

The fluid flow time-averaged over the whole duration $t_0$ of the movie, $\left<\bm{u}(r,\theta)\right>_t = \frac{1}{t_0} \int_0^{t_0} \bm{u}(r,\theta,t)\,dt$ (see Fig.~\ref{fig1}(c)), is described accurately by the modes of the `squirmer' expansion of the flow field around a sphere \cite{Lighthill:1952,Blake:Squirmer,Magar:2003} modified to take into account the net force exerted by the holding pipette. The results are in agreement with previous measurements on freely swimming colonies \cite{Drescher:2010kx}, and support the hypothesis that flagellar dynamics is the same in held and freely swimming colonies, a fact already established  for the closely related unicellular species {\it Chlamydomonas} \cite{Polin:2009kx}.
Subtracting the time average from the instantaneous flow field, $\bar{\bm{u}} = \bm{u} - \left<\bm{u}\right>_t$, 
highlights the flow's dependence on the local flagellar phase within the beating cycle. It is this phase that we wish to measure.
Figure~\ref{fig2} shows representative kymographs, through the same time interval, for the radial and 
tangential components of $\bm{u}$ at $r^{*} = 1.3\,R$. This value of $r^*$, which corresponds to the dashed circle 
in Fig.~\ref{fig1}(a), has been found empirically to maximise the kymographs' signals in 
most experiments. 

Temporal variations in the flow are different for the two components (Figs.~\ref{fig2}(a,b)):
$\bar{u}_{r}$ exhibits a wave travelling from the anterior to the posterior poles of the colony (see also Fig.~\ref{fig1}(d)); $\bar{u}_{\theta}$ is 
closer to a standing wave.
Such discrepancy should in fact be expected for a MW propagating along a spherical distribution of flagella as found 
in {\it Volvox}: $\bar{u}_{\theta}$ is dominated by the flow induced by the flagella in the equatorial region, which 
are (approximately) mutually synchronized and are much more numerous than those at other latitudes in the colony. 
Peaks in $\bar{u}_{\theta}$ would then happen in the middle of the power stroke of the equatorial flagella, and 
should  correspond to minima in the magnitude of $\bar{u}_{r}$ around the equatorial region: this is borne out by 
the experimental kymographs.
We confirmed this hypothesis both by visual inspection of flagellar behaviour and by simulations of arrays of model 
oscillators (see below). The simulations show clearly that only the component of the velocity perpendicular to 
the no-slip surface (for us $\bar{u}_{r}$) tracks closely the local phase of the oscillators, provided that the phase profile 
changes sufficiently slowly, as is the case for {\it Volvox}. 
This finding is also consistent with the fact that the flow induced by a force normal to a no-slip plane decays 
faster -- and is then more localised -- than that from an equivalent tangential force ($r^{-4}$ vs. $r^{-3}$ at constant 
distance from the surface). $\bar{u}_r(r^*,\theta,t)$ can then be used as a proxy for the local flagellar phase 
along the beating cycle, eliminating the need to track the motion of individual flagella on the colony's surface. 
Figure~\ref{fig2}(a) corresponds to a MW which propagates in the direction of the flagellar power stroke, 
called symplectic, and is representative of all the 60 colonies examined (see full experimental dataset for 
additional results). Direct inspection of several colonies at different orientations confirmed that {\it Volvox} 
MWs have only a minimal lateral (diaplectic) component.
The average properties of {\it Volvox} MWs can be quantified \cite{Brumley:2012,Elgeti:2013} through the 
normalised autocorrelation function of $\bar{u}_r(r^*,\theta,t)$ (Fig.~\ref{fig2}(c)), 
$C(\Delta\theta, \Delta t) = \left< {\bar u}_r(r^*,\theta+\Delta\theta, t+\Delta t)\,{\bar u}_r(r^*,\theta,t)\right>/
\left<{\bar u}_r(r^*,\theta,t)^2 \right>$ where  $\left<\right>$ denotes averaging over all $t$ and $\theta$. 
The phenomenological functional form $C_{\text{fit}}(\Delta \theta, \Delta t) = \exp{\left(-\frac{\Delta t}{\tau}\right)}
\exp{\left(-\frac{\Delta\theta}{L}\right)} \cos(k\,\Delta\theta - \omega\,\Delta t)$ fits the experimental autocorrelation function 
remarkably well, with average deviation of $0.033 \pm 0.011$ between the two, across all colonies observed.  
From these fits we estimate the mean beating frequency  $f= \omega/2\pi = 1/T =  32.3 \pm 3.1 \ $Hz; the 
autocorrelation decay time $\tau/T = 3.4 \pm 2.0 $, and decay length $L=0.35 \pm 0.10$; and the 
mean wavenumber within the population $k= 4.7 \pm 0.9$ (twice the number of complete waves along a meridian). 
$k$ is  positive in all our experiments, indicating the ubiquitous presence of a symplectic MW which emerges despite 
the absence, in this species, of any direct intercellular connection between somatic cells. 

By averaging MW properties, we necessarily mask brief inhomogeneities in the beating dynamics. These are primarily 
not random, but rather appear as recurrent defects in the MW pattern, as circled in Fig.~\ref{fig2}(a). 
To our knowledge, such defects have never previously been observed experimentally, although a  similar behaviour 
was reported in recent simulations \cite{Elgeti:2013}.
Reminiscent of so-called frequency plateaus observed in strips of coupled oscillators \cite{Ermentrout1986}, the defects 
amount to global slip events, where the phase difference between two groups of oscillators suddenly increases by a full 
cycle \cite{Goldstein:2009vn,Goldstein:2011_emergence,Polin:2009kx,Brumley2014}. In the majority of our experiments, 
the defects are directly recognisable from the kymograph (see Fig.~\ref{fig2}(a)). They consistently 
appear around the equatorial region, and correspond always to posterior flagella slipping ahead of anterior 
ones. This results in loss of temporal and spatial correlation of the MW, and it is responsible for the surprisingly 
short correlation time ($\tau/T$) and length ($L$) observed. 
The presence of defects results in the $\theta$-dependent average beating frequency seen in Fig.~\ref{fig2}(g), which could reflect a combination of factors including 
a physiological bias in the intrinsic flagellar beating frequency across the colony, or the effect of lower drag due to higher concentration of somatic cells in the colony's posterior. 
Inspection of the power spectrum of the autocorrelation function of the kymograph signal at {\it individual} values of $\theta$, reveals that this transition in average beating frequency has a characteristic structure. Close to the top or to the bottom of the colony, all cells are phase-locked and beat with a single frequency, noticeably different for the two groups (the main peaks in the red and blue curves in Fig.~\ref{fig2}(f); the peaks' width is caused by both intrinsic beating variability and measurement error). 
The transition region, where defects appear, interpolates between these two groups by combining their frequencies. The result is a double peak in the power spectrum (e.g. green double peak in Fig.~\ref{fig2}(f)), where the relative height of the peaks  depends on the proximity to one or the other of the phase-locked groups.  This phenomenology is characteristic of all the colonies displaying clearly defined phase defects.

Notice that the frequency bias would by itself generate a wave propagating in the direction {\it opposite} to the experimental one, and hence cannot be the 
fundamental mechanism selecting the direction of the observed MW. In what follows we discuss the behaviour of 
a minimal model of interacting rotors which spontaneously generates a symplectic MW, and the effect on these dynamics of a large scale bias as found in  Fig.~\ref{fig2}(g).
 

\section*{Hydrodynamic model for interacting flagella} 

\begin{figure}[t!]
\begin{center}
\includegraphics[width=0.93\columnwidth]{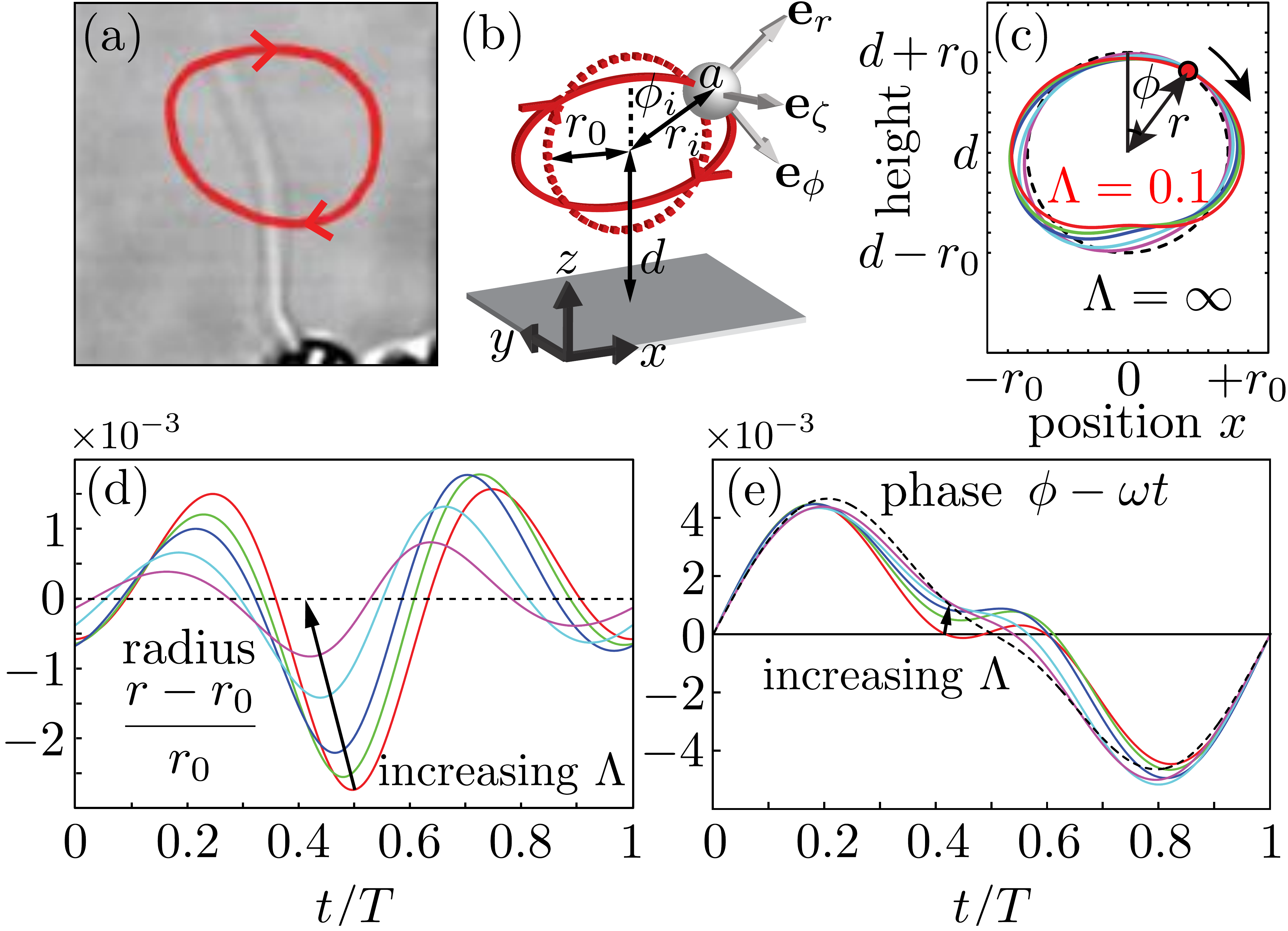}
\caption{Modelling flagella. (a) Tip trajectory over $10$ beats of a flagellum of a {\it Volvox} somatic cell.  
(b) Rotor as a model flagellum: a sphere of radius $a$ elastically bound to a circular trajectory of radius $r_0$ (dashed red) 
perpendicular to a no-slip plane, driven by a tangential force in the $\phi$-direction. (c) Intrinsic trajectory 
of an isolated rotor. Steady state limit cycle of the sphere above the no-slip wall at $z=0$, for 
$\Lambda = 0.1,1,2,5,10,\infty$. Perturbations from the circular trajectory have been magnified by a factor of 
100 so that the shape is clearly visible. The evolution of the (d) radius and (e) geometric phase through 
one beating period are also shown.}
\label{fig3}
\end{center}
\end{figure}

The surface-mounted somatic cells of {\it Volvox} are a few tens of microns apart, and their flagella are therefore more 
nearly in the weak-coupling limit than the cilia of historically studied organisms such as {\it Paramecium}. It is thus appropriate to model the fluid disturbance produced by their operation as a multipole expansion \cite{Vilfan:2012}, of which we will only keep the mode with the slowest spatial decay -- the Stokeslet -- representing the effect of a point force. 
This flow is analogous to the far field of a rigid sphere pulled through the fluid.
It has recently been shown that representing a {\it Volvox} flagellum as a single Stokeslet provides an accurate representation of its flow field down to distances of $\sim10\,\mu$m, smaller than those typically separating cells within colonies ($\sim20\,\mu$m) \cite{Brumley2014}. High-speed tracking of the flagellar waveform combined with resistive force theory also confirms that the distributed forces associated with the motion can be well represented by a single point force which periodically traverses a closed loop \cite{Brumley2014}.
Inspired by the trajectories of flagellar tips in {\it Volvox}, 
and following an approach similar to others \cite{Uchida:2011kx, Niedermayer:2008fk, Brumley:2012}, a beating flagellum 
will thus be modelled as a small sphere of radius $a$ elastically bound to a circular trajectory of 
radius $r_0$ by radial and transversal springs of stiffnesses $\lambda$ and $\eta$ respectively, and 
driven by a tangential force of magnitude $f^{\text{drive}}$ (see Fig.~\ref{fig3}). The position 
of the sphere is given by $\bm{x} = \bm{x}^0 + \bm{s} (\zeta,r, \phi)$, where 
$\bm{s} = (r \sin(\phi),\zeta, r \cos(\phi))$. The 
prescribed trajectory, defined by $(\zeta=0, r=r_0)$, is perpendicular to a no-slip plane at 
$z=0$ representing the surface of {\it Volvox}, and its centre $\bm{x}^0$ is at a distance $d$ from the plane. 
The proximity of the no-slip boundary causes an asymmetry in the sphere's motion which induces a net flow, thus 
mimicking  power and recovery strokes of real flagella.

The sphere's velocity $\bm{v}$ follows the force balance requirement of Stokes flow: 
${\bm \gamma}({\bm x}) \cdot \bm{v} = -\lambda (r - r_0)\,\bm{e}_{r} - \eta\zeta\,\bm{e}_{\zeta} 
+ f^{\text{drive}}\bm{e}_{\phi}$. Here ${\bm \gamma} = {\bm \gamma} (\bm{x}) = \gamma_0 \big[ \textbf{I} + 
(9a/16z) \big(\textbf{I} + \bm{e}_z \bm{e}_z \big) + \mathcal{O}((a/z)^3) \big]$ is the friction tensor associated 
with motion near the no-slip wall \cite{Dufresne2000}, and $\gamma_0 = 6 \pi \mu a$ is the drag on a 
sphere of radius $a$ in an unbounded fluid of viscosity $\mu$.
Because the drag is not isotropic, the trajectory of an isolated sphere will deviate slightly from the 
prescribed circle ($\delta r/r_0 \sim 0.5 \%$ here) to a new limit cycle $(r(t),\phi(t),\zeta(t))$ of period $T$, 
which we  obtain directly from the simulations for each set of parameters. This limit cycle will represent the 
unperturbed state of the single rotor. 
Figure~\ref{fig3} shows the polar coordinates $(r(t),\phi(t))$ of such limit cycles, for various 
values of the dimensionless ratio $\Lambda = \lambda d / f^{\text{drive}}$. This parameter governs the 
deformability of the rotors' 
orbit:  larger (smaller) values of $\Lambda$ mean stiffer (softer) radial recoil compared to the driving force, 
and result in smaller (larger) deviations from $r=r_0$.
For {\it Volvox} flagella, with bending rigidity $\kappa = 4 \times 10^{-22}\,\textrm{N}\,\textrm{m}^2$
\cite{Niedermayer:2008fk} and length $L \sim 10\,\mu$m, small deflections of the filament would be resisted 
with an effective spring constant $\lambda \simeq \kappa / L^3 = 4 \times 10^{-7}\,\textrm{N}\,\textrm{m}^{-1}$.
Approximating the driving force as $f^{\text{drive}} = 2 \pi r_0 \gamma_0/T$, and taking 
$a \sim 1\,\mu$m, $T\sim 1/33\,$s and $d/r_0 \sim 1$ \cite{Brumley:2012}, the normalised spring 
constant is estimated as $\Lambda \sim 0.1$.

From the limit cycle, we define the phase of the rotor, $\Phi(\phi)$, such that $\dot{\Phi} = \omega = 2 \pi / T$
\cite{Pikovsky:2003}. The presence of the wall implies that the phase $\Phi$ is different from the angle $\phi$ even 
for completely rigid orbits. The phase will be useful when studying the dynamics of coupled rotors, 
since any perturbations from constant evolution of $\Phi$ will be solely due to hydrodynamic interactions between 
different rotors.
For a system of $N$ rotors, the net hydrodynamic force acting on the $i^{\text{th}}$ rotor is
\begin{equation}
\bm{F}_i = - {\bm \gamma}_i \cdot \bigg[  \bm{v}_i +  
\sum_{j \neq i} \textbf{G}(\bm{x}_j,\bm{x}_i) \cdot \bm{F}_j \bigg], \label{F_i_N_spheres}
\end{equation}
where $\textbf{G}(\bm{x}_j,\bm{x}_i)$ is the Green's function that couples the different spheres in the 
presence of the no-slip wall \cite{Lorentz:1896,Blake:1971uq}.  The time evolution of the system of $N$ moving 
spheres is then given by 
\begin{equation}
\left[\textbf{R}^{T} \cdot \textbf{M}^{-1} \cdot \bm{\Gamma} \cdot \textbf{R}\right]_{ji} \left( \begin{array}{c}
r_i\dot{\phi}_i \\
\dot{r}_i \\
\dot{\zeta}_i \\
\end{array} \right) = \left( \begin{array}{c}
 f^{\text{drive}} \\
-\lambda\left(r_j - r_0\right) \\
- \eta\,\zeta_j \\
\end{array} \right). 
\label{matrix_system}
\end{equation}
The $3N \times 3N$ matrices $\bm{\Gamma}$, $\textbf{M}$ and $\textbf{R}$ are defined in terms of their constitutive 
$3 \times 3$ blocks  $\bm{\Gamma}_{ij} = \delta_{ij} {\bm \gamma}(\bm{x}_i)$, $\textbf{M}_{ij} = 
\delta_{ij} \textbf{I} + \big( 1 - \delta_{ij} \big) {\bm \gamma}(\bm{x}_i) \cdot \textbf{G}(\bm{x}_j,\bm{x}_i)$, and 
$\textbf{R}_{ij} = \delta_{ij}\textbf{R}_i$, where the columns of the rotation $\textbf{R}_i$ are given by the vectors 
$\bm{e}_{\phi_i}$, $\bm{e}_{r_i}$ and $\bm{e}_{\zeta_i}$ respectively \cite{Brumley:2012}.

\section*{Pair Synchronization}

\begin{figure}[t!]
\begin{center}
\includegraphics[width=\columnwidth]{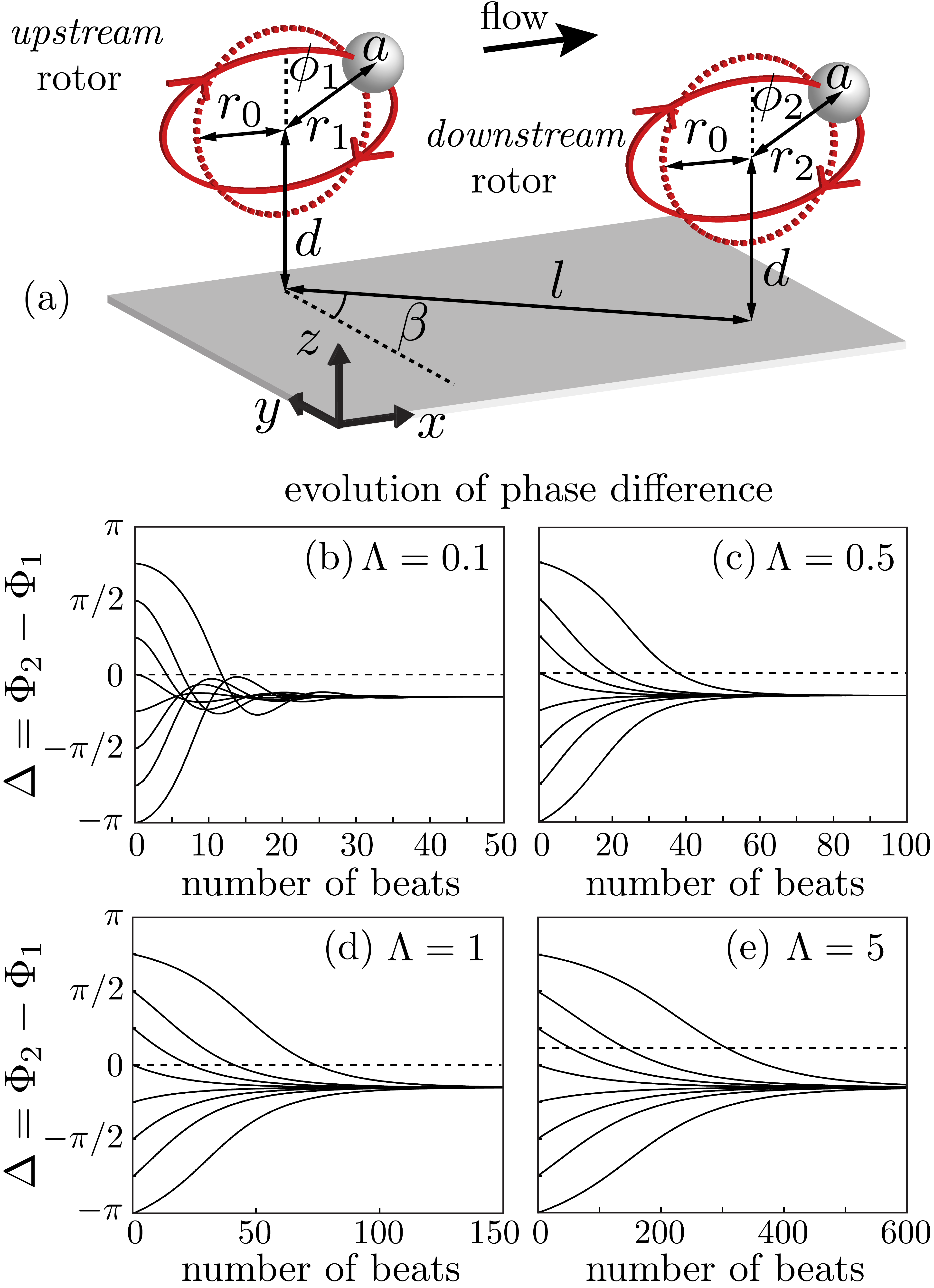}
\caption{Synchronization dynamics of pairs of rotors. (a) Schematic diagram of a pair of interacting rotors, each of which has three 
spatial degrees of freedom; geometric phase $\phi_i$, trajectory radius $r_i$ and out-of-plane displacement $\zeta_i$ 
(not shown). The rotor-rotor spacing $l$ and planar angle $\beta$ are depicted. (b--e) Simulations of interacting pairs 
of rotors, for various spring stiffness $\Lambda = \lambda d / f^{\text{drive}}$. In each case, the system 
converges to a steady state phase difference with $\Delta < 0$, during a time interval that increases with $\Lambda$. 
The parameters used are $a/d=0.01$, $r_0/d=0.5$,  $\beta = \pi/2$, $l/d=2$.}
\label{fig4}
\end{center}
\end{figure}

Pair synchronization is the building block of large scale coordination, and it will be studied here for two identical rotors spaced a distance $l$ apart, at an angle $\beta$ in the $x$-$y$ plane (Fig.~\ref{fig4}(a)). 
For sufficiently stiff springs, hydrodynamic interactions induce only small perturbations around the rotors' limit cycles, 
and the whole system can then be described just in terms of the two phases $(\Phi_1,\Phi_2)$.  
Combining the phases into their sum, $\Sigma = \Phi_2+\Phi_1$, and difference, $\Delta =  \Phi_2-\Phi_1$, provides 
a  convenient description of the synchronization state of the system, which is simply characterised by $\Delta(t)$.  
To develop intuition, we work almost exclusively within this phase-oscillator regime, although simulations 
use the full dynamical system of Eq.~\eqref{matrix_system}.

Figures~\ref{fig4}(b-e) show the evolution of two coupled spheres oriented so that one is directly
downstream of the other ($\beta = \pi/2$), for four different values of $\Lambda$. 
In every case, the rotors are able to achieve a phase-locked state by perturbing one another from their limit cycles, 
and the system converges to a steady state in which $\Delta = \Delta_S$, independently of the initial condition. The fact 
that $\Delta_S<0$ means that the rotor downstream lags behind the one upstream, a condition which would
be naturally expected to produce a symplectic metachronal wave along a strip of rotors. We will see below 
that this intuition is not necessarily correct. \\

\begin{figure*}[t!]
\begin{center}
\includegraphics[width=\textwidth]{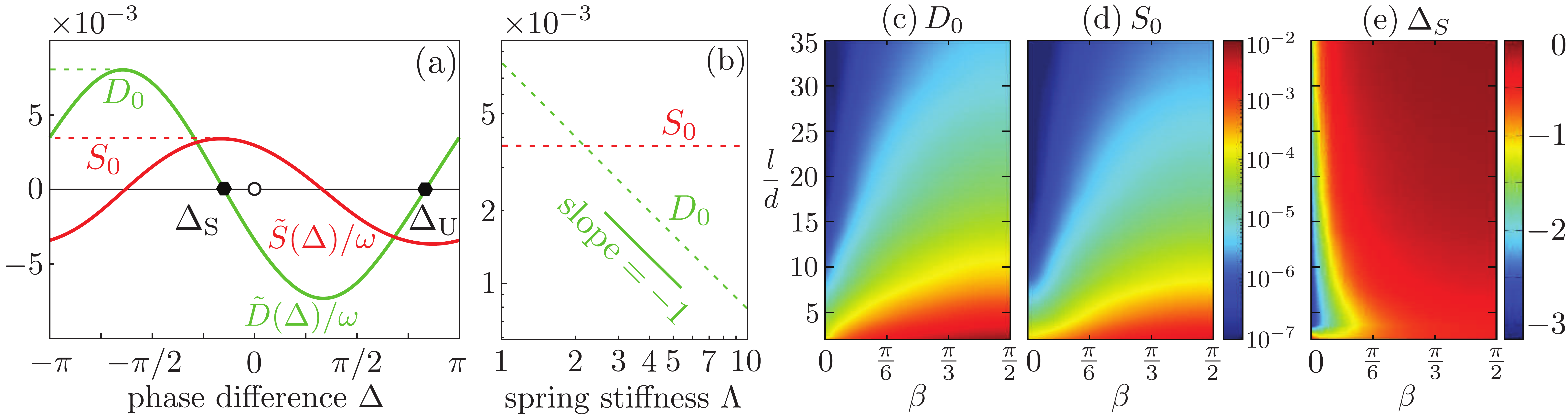}
\caption{Phase oscillator reduction. (a) Coupling functions $\widetilde{D}(\Delta)$ and 
$\widetilde{S}(\Delta)$ for a pair of interacting spheres. The corresponding parameters are $a/d = 0.01$, $r_0/d = 0.5$, $\beta = \pi/2$, $l/d=2$, $\Lambda = \lambda d / f^{\text{drive}} = 1$. (b) Amplitude of coupling functions for different 
spring stiffnesses. Computed amplitudes of (c) $\widetilde{D}(\Delta)/\omega$ and (d) $\widetilde{S}(\Delta)/\omega$, 
as well as (e) the stable equilibrium phase difference $\Delta_S$, shown as functions of rotor-rotor spacing 
$2 \leq l/d \leq 35$ and offset angle $0 \leq \beta \leq \pi/2$.}
\label{fig5}
\end{center}
\end{figure*}

While $\Delta_S$ depends very little on the precise value of $\Lambda$, this is not the case for the timescale for 
phase-locking, which is seen to increase for progressively stiffer rotors. To understand the origin of this dependence, 
we write the evolution of the phase difference $\dot{\Delta}= \omega^- + D(\Delta,\Sigma)$ and sum $\dot{\Sigma} 
= \omega^+ + S(\Delta,\Sigma)$ in terms of effective couplings $D$ and $S$, where $\omega^{\pm} = 
\omega_2 \pm \omega_1$.
When the phase difference evolves much more slowly than the sum, the equations can 
be averaged over the characteristic timescale of $\Sigma$ to obtain the time-averaged system:
\begin{eqnarray}
\dot{\Delta} = \omega^{-} + \widetilde{D}(\Delta) \label{Delta_dot_averaged}\\
\dot{\Sigma} = \omega^{+} + \widetilde{S}(\Delta)\label{Sigma_dot_averaged}
\end{eqnarray}
The crossing of curves in Fig.~\ref{fig4}(b) indicates that for that value of $\Lambda$ the 
evolution of the system is not uniquely determined by $\Delta$: for these parameters the time-averaged equations 
do not provide a good approximation of the dynamics. Conversely, each panel in Figs.~\ref{fig4}(c-e) contains a set of curves that, from a given value of $\Delta$, approach the asymptote in essentially the same manner, supporting the system's description in terms of the single variable $\Delta$.
Figure~\ref{fig5}(a) shows the coupling functions $\widetilde{D}(\Delta)$ and 
$\widetilde{S}(\Delta)$ determined numerically from simulations for one set of parameters, and nondimensionalised 
by the average beat frequency $\omega$ ($\omega = \omega_1 = \omega_2$ in this case). 
These curves are characteristic of the whole parameter range we have studied ($0.1 < \Lambda < 10$).  To great
accuracy they are, respectively, $-D_0\sin(\Delta-\Delta_S)$ and $S_0\cos(\Delta-\Delta_S)$. 
The system has  a stable fixed point at  $\Delta=\Delta_S<0$, which also corresponds to the maximum 
in $\widetilde{S}(\Delta)$: when phase-locked, the rotors increase their average beat frequency owing to a lower 
effective drag \cite{Reichert:2005qf,Niedermayer:2008fk}. Apart from an increase in speed, 
this cooperative drag reduction does not influence 
the synchronization of a single pair, but it will have a major impact on the collective state of larger groups of rotors. 
Figure~\ref{fig5}(b) shows the dependence of the amplitudes $D_0$ and $S_0$ on the effective 
stiffness $\Lambda$. While $S_0$ is independent of $\Lambda$, reflecting the fact that drag reduction is a 
purely hydrodynamic effect, the amplitude $D_0$, which governs the convergence towards the stable fixed point, 
scales as $D_0 \sim \Lambda^{-1}$ (Fig.~\ref{fig5}(b)). This is reflected in the 
$\Lambda$-dependence of the rate of convergence towards phase-locking seen in Fig.~\ref{fig4}(b).
{\it Stiff} ($D_0 < S_0$) and {\it soft} ($D_0 > S_0$) regimes can be recognised, and the position of the system with 
respect to this divide can be tuned simply by adjusting the parameter $\Lambda$, notably by changing the radial 
stiffness of the orbits. In the asymptotic limit as $\Lambda \rightarrow \infty$, phase-locking does not occur 
(for these parameters) -- a phenomenon represented by $D_0 \rightarrow 0$. As reported above, we also find 
that $\Delta_S$ is independent of $\Lambda$.

The two most significant parameters governing the spatial configuration of the pair of rotors are the 
rotor-rotor spacing $l$ and the angle $\beta$ (see Fig.~\ref{fig4}(a)). 
Figures~\ref{fig5}(c-d) show the dependence of the coupling strengths $(D_0,S_0)$ on 
$(l,\beta)$: both decrease  
$ \sim l^{-3}$ with increasing $l$, while at fixed separation they increase monotonically as $\beta$ goes from $0$ 
(side-by-side) to $\pi/2$ (aligned downstream). This behaviour follows semi-quantitatively from the (average) magnitude 
of the flow generated by a single rotor, which in the far field is proportional to $\sin(\beta)/l^3$ \cite{Blake:1971uq}.
$\Delta_S$, which is consistently negative, shows instead a much weaker dependence across almost all of the region 
of parameter space explored (see Fig.~\ref{fig5}(e)). 
It decays slowly with increasing $l$ ($\sim l^{-1}$ for $l/d>10$ and $\beta\sim\pi/2$) and is fairly insensitive 
to changes in $\beta$, except for a narrow region around $\beta=0$ showing strong dependence on the position 
of the two rotors ($\Delta_S$ must be an odd function of $\beta$). 
The phase lag at synchronization is for the most part almost unaffected by the detailed configuration of the two 
rotors, in sharp contrast to previous models \cite{Vilfan:2006uq, Niedermayer:2008fk}.
Simulations for a range of other physical parameters, including sphere size and trajectory radius, show 
qualitatively identical results.

{\bf Influence of a variable driving force.---} 
Throughout its beating cycle, it is plausible that a flagellum will exert a changing force on the surrounding fluid 
and this variation, as opposed to waveform compliance, has been proposed as a major factor leading to phase locking \cite{Uchida:2011kx}.
\begin{figure*}[t!]
\begin{center}
\includegraphics[width=\textwidth]{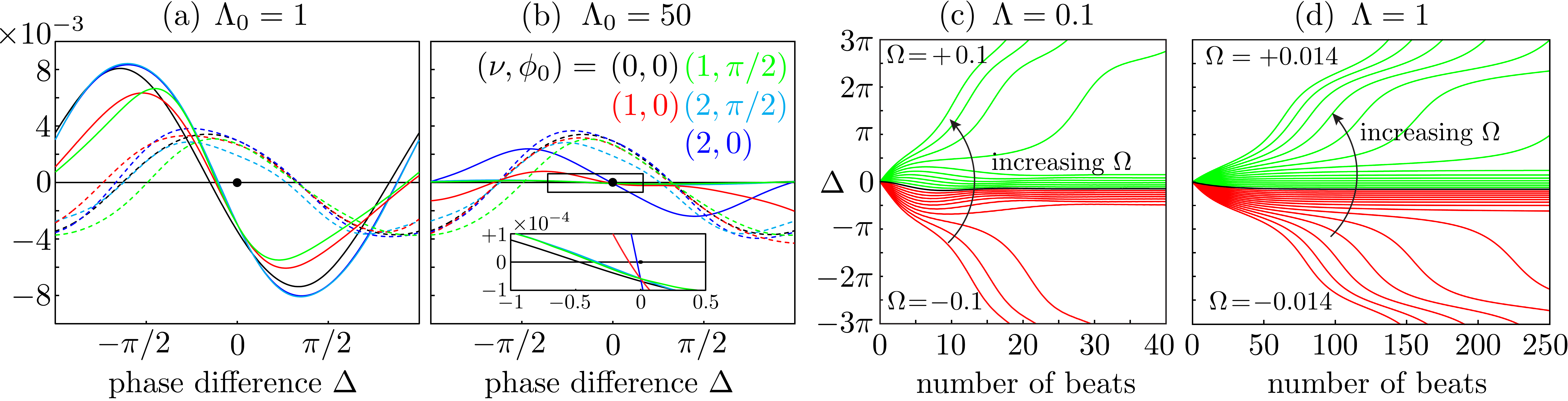}
\caption{Nondimensionalised coupling functions $\widetilde{D}(\Delta) / \omega$ (solid) and 
$\widetilde{S}(\Delta)/ \omega $ (dotted) for a pair of interacting spheres subject to the driving force $f^{\text{drive}} 
\sim 1 + 0.5 \sin (\nu \phi + \phi_0)$ for various values $(\nu,\phi_0)$. Plots are shown with (a) $\Lambda_0  = 1$ and 
(b) $\Lambda_0  = 50$. Detuning the rotors: evolution of the phase difference $\Delta$ for (c) $\Lambda = 0.1$ and 
(d) $\Lambda = 1$. Results are shown for various values of $\Omega = 2\omega^{-} / \omega^{+}$. Positive and 
negative values of $\Omega$ are shown in green and red respectively, while the black curves represent 
$\Omega = 0$ (identical rotors). All figures correspond to the parameters $a/d = 0.01$, $r_0/d = 0.5$, $\beta = \pi/2$,
$l/d=2$.}
\label{fig6}
\end{center}
\end{figure*}
Indeed, force modulation can by itself establish synchronization in pairs of model oscillators with rigid 
trajectories \cite{Uchida:2011kx}, and recent studies \cite{Kotar2013} have shown that in specific circumstances 
this can dominate over synchronization mediated by orbit compliance.
Here we consider the effect of a modulated driving force $f^{\text{drive}} = f_0(1 + C \sin (\nu \phi + \phi_0))$ in 
our system. The average  amplitude $f_0$ is used to define the nondimensional stiffness $\Lambda_0 = \lambda d/f_0$. 
Figure~\ref{fig6} shows the coupling functions $\widetilde{D}(\Delta)$ 
and $\widetilde{S}(\Delta)$ for $C=0.5$, $\nu \in \{0,1,2\}$, and $\phi_0 \in \{0,\pi/2\}$. Notice that $\nu=2$ 
corresponds to a class of functional forms highly effective in establishing synchronization in circular 
rotors \cite{Uchida:2011kx,Kotar2013}.
For $\Lambda_0 = 1$ (and smaller), modulating the driving force has no significant bearing on the results. 
While changes in $D_0$ can be significant ($\lesssim30\%$; but $D_0>S_0$ always), the difference is mainly 
far from the phase-locked state, and both $\Delta_S$ and $S_0$ do not vary appreciably from their values at constant 
driving force. In this regime synchronization is dominated by orbit compliance. The corresponding coupling functions for 
$\Lambda_0 = 50$, approximating the rigid case, are presented in Fig.~\ref{fig6}(b).
$S_0$ is essentially unchanged, but the phase-locking dynamics is severely altered. Both $D_0$ and  $|\Delta_S|$ have 
been drastically reduced. The most robust synchronization appears for $(\nu,\phi_0) = (2,0)$, with a $75\%$ reduction 
in $D_0$, and this is accompanied by a lag of just $|\Delta_S| \sim 0.03$  (see
Fig.~\ref{fig6}(b) inset). As we will see, a system with these characteristics does not generate a MW in a finite chain of rotors, an 
essential requirement for any realistic model of metachronism. Of course, there may be specific fixed trajectory shapes 
and force profiles that give rise to an appropriate coupling. However, orbit compliance is capable of facilitating rapid 
phase-locking in a manner essentially independent of the driving force, in particular for $\Lambda_0\sim0.1$. 
Additional simulations were undertaken with varying rotor geometries, including changes in $\beta$ and $l$, 
but the results are qualitatively unchanged. 

{\bf Detuning the beating frequencies.---}Even within the same group of cells, different flagella will generally have 
slightly different frequencies. Figure~\ref{fig2}(g) shows that this is the case for {\it Volvox} colonies, with 
a $\sim\! 10\%$ spread in (apparent) beating frequencies. Here we study the impact of such inhomogeneity 
on synchronization of a pair of rotors. We focus on frequency differences caused by variations in driving force.  
The difference between the rotors will be characterised by the fractional frequency difference 
$\Omega = 2\omega^-/\omega^+=(\omega_2-\omega_1)/\bar\omega$, where $\bar\omega=(\omega_1+\omega_2)/2$
is the mean frequency.
Figure~\ref{fig6}(c,d) shows the evolution of the phase difference $\Delta$ for a 
pair of rotors, beginning at $\Delta=0$ for $t=0$, for various values of $\Omega$. The curves have been computed 
for $\Lambda =0.1$ and $1$, where $\Lambda$ is defined in terms of the average driving force. The parameters
corresponding to the smallest $\Lambda$ provide the closest approximation to real {\it Volvox} cells within this type 
of model.

\indent Phase-locking is clearly possible for $\Omega$ in a finite interval around zero, albeit at different values of 
$\Delta_S$.  Figure~\ref{fig6}(c) demonstrates that stable phase-locking can 
occur for $\Lambda = 0.1$ even when the rotors possess a $|\Omega| \sim 6\%$ difference in their intrinsic 
beat frequencies, which compares well with the measured value of $10\%$ for flagella isolated from {\it Volvox} \cite{Brumley2014}
However, for the larger value of $\Lambda = 1$, the phase-locking is less robust, and the system 
can support a maximum detuning of only $\sim 0.7\%$ before drifting indefinitely.
As $\widetilde{D}(\Delta) \simeq -D_0 \sin(\Delta - \Delta_S)$, the overall evolution of the phase difference 
can then be written as $\dot{\Delta} = \Omega - D_0 \sin (\Delta - \Delta_S^0)$ (time is rescaled using the average 
beating frequency). Stationary points occur for $\Delta_S(\Omega) = \Delta_S^0 + \arcsin( \Omega / D_0 )$, which 
has a solution only for $|\Omega| \leq D_0$. Because $D_0 \sim \Lambda^{-1}$, phase-locking will occur below a 
threshold mismatch $|\Omega|_{\text{max}} \sim \Lambda^{-1}$. This estimate, based on the time-averaged 
phase dynamics, is supported by the full hydrodynamic simulations, which show that an order of magnitude 
increase in $\Lambda$ (from 0.1 to 1) corresponds to an order of magnitude decrease in the threshold 
detuning (from $\sim\! 6\%$ to $\sim\! 0.7\%$).

\begin{figure*}[t!]
\begin{center}
\includegraphics[width=\textwidth]{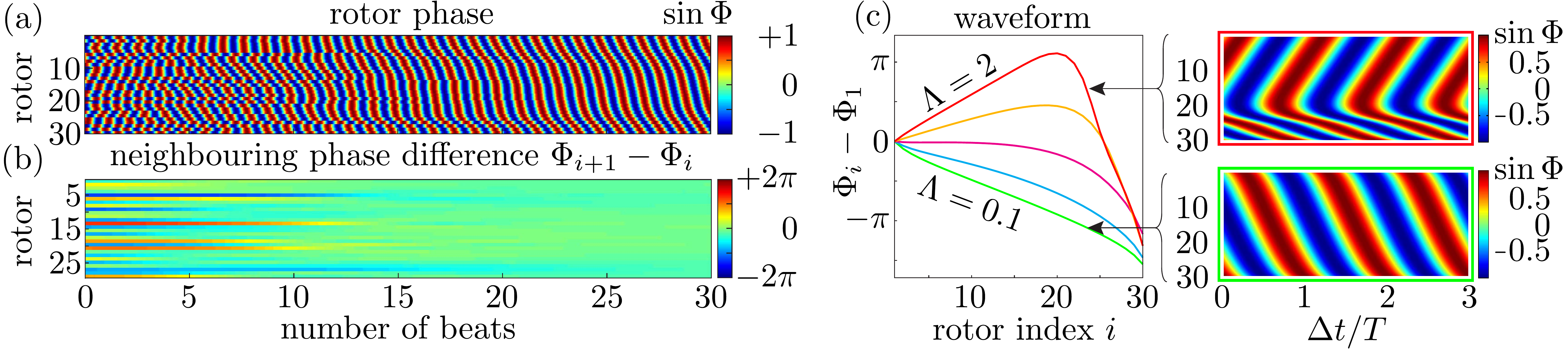}
\caption{Metachronal wave development. (a) The phase along the chain of rotors 
and (b) the neighbouring phase difference $\Phi_{i+1} - \Phi_{i}$ are shown as functions of time. The symplectic 
MW is established within 15 beats for this value of $\Lambda=0.1$. (c) Phase profile of an array of 
$30$ spheres after $t/T=1200$ beats. Results are shown for 
$\Lambda = \lambda d/f^{\text{drive}} = 0.1,0.5, 1, 1.5, 2$. All results correspond to 
parameters $a/d = 0.01$, $r_0/d = 0.5$, $l/d=2$.}
\label{fig7}
\end{center}
\end{figure*}

\section*{Group Synchronization}

{\bf Constant driving force.---}Although necessary, pair synchronization is not a sufficient condition to guarantee that a 
group of rotors will generate MWs. This fact is evidenced by studying the behaviour of a finite linear chain of 
rotors, a configuration in which any realistic model of metachronal coordination should be able to produce robust MWs. 
We describe here the behaviour of linear chains of 30 identical rotors, aligned in the downstream direction, 
with $(a/d,r_0/d,l/d) = (0.01,0.5,2)$. These are meant to mimic the behaviour of cells along a single meridian of 
{\it Volvox}, if they could be observed in isolation. The effective stiffness  $\Lambda$ will be varied from 0.1 
({\it Volvox}-like) to 2. Similar studies were performed for chains of length ranging from 3 to 50 rotors, 
and $2 \leq l/d \leq 20$: the qualitative features of the dynamics remain the same. 

Figure~\ref{fig7} shows that for $\Lambda=0.1$ a steady symplectic MW develops from random initial 
conditions within $\sim\!15$ beats. 
As in a single pair, this phase-locking timescale is stiffness dependent, and within our range of $\Lambda$ it varies 
from tens to hundreds of beating cycles, independently of initial conditions, eventually diverging for 
$\Lambda \rightarrow \infty$. Figure~\ref{fig7}(c) shows the phase profiles along the chains after 1200 beats, representative of the 
general behaviour. We see that the symplectic MW, defined as $\Phi_{i+1}<\Phi_{i}\; \forall \ i$, morphs into a chevron 
pattern as the rotors' stiffness is increased. For this configuration, $\Lambda\simeq1$ marks the boundary between the 
two different patterns. Phase locking into a chevron has been previously reported for similar linear systems \cite{Sakaguchi:1988, Niedermayer:2008fk} which were not able, however, to generate also a stable 
metachronal coordination.

Qualitatively, chevron formation arises from the competition between drag reduction at synchronization and the tendency 
of neighbouring rotors to phase-lock at a given $\Delta_S\neq0$. Drag reduction is more pronounced for rotors in 
the chain's interior, which will tend to lead those at the extremes. This would result in a large fraction of the chain having 
a phase profile opposite to the one required by metachronal coordination. In the stiff regime $(D_0 < S_0)$, drag 
reduction dominates, and leads to a chevron-like steady state phase profile. However, for sufficiently compliant 
trajectories $(D_0 > S_0)$ the drive towards locking at a given phase lag is strong enough to overcome the effect 
of drag reduction, and nearest neighbours will synchronize with lags of the same sign as $\Delta_S$. In the present system 
this results in a symplectic MW. 
In fact, even within the MW regime, long range hydrodynamic interactions between rotors at all separations 
$l$ will contribute to determine the steady-state phase lags between nearest neighbours. Because 
$\Delta_S\sim l^{-1}$ and $D_0\sim l^{-3}$ we can generally expect the magnitude of this phase lag to be smaller 
than for an isolated pair at the equivalent separation. For example, while  a pair with $\Lambda = 0.1$ would phase-lock 
with $\Delta_S = -0.47$ (other parameters as in Fig.~\ref{fig7}), the average nearest neighbour phase lag
that is actually realised in a linear array is significantly less, $\bar{\Delta}_S = -0.17$. The nonlocal interactions are 
clearly important in determining the overall system dynamics. With these parameters, an array of 30 rotors should 
provide the best approximation to a single meridian of {\it Volvox} cells from a real colony. Our simulations predict 
that the system should develop a symplectic MW, indeed observed experimentally, with a wavenumber 
$k \simeq 2$. Given the simplicity of the model, we believe that this is in good agreement with the experimental 
average value of $k= 4.7 \pm 0.9$.

\begin{figure*}[t!]
\begin{center}
\includegraphics[width=\textwidth]{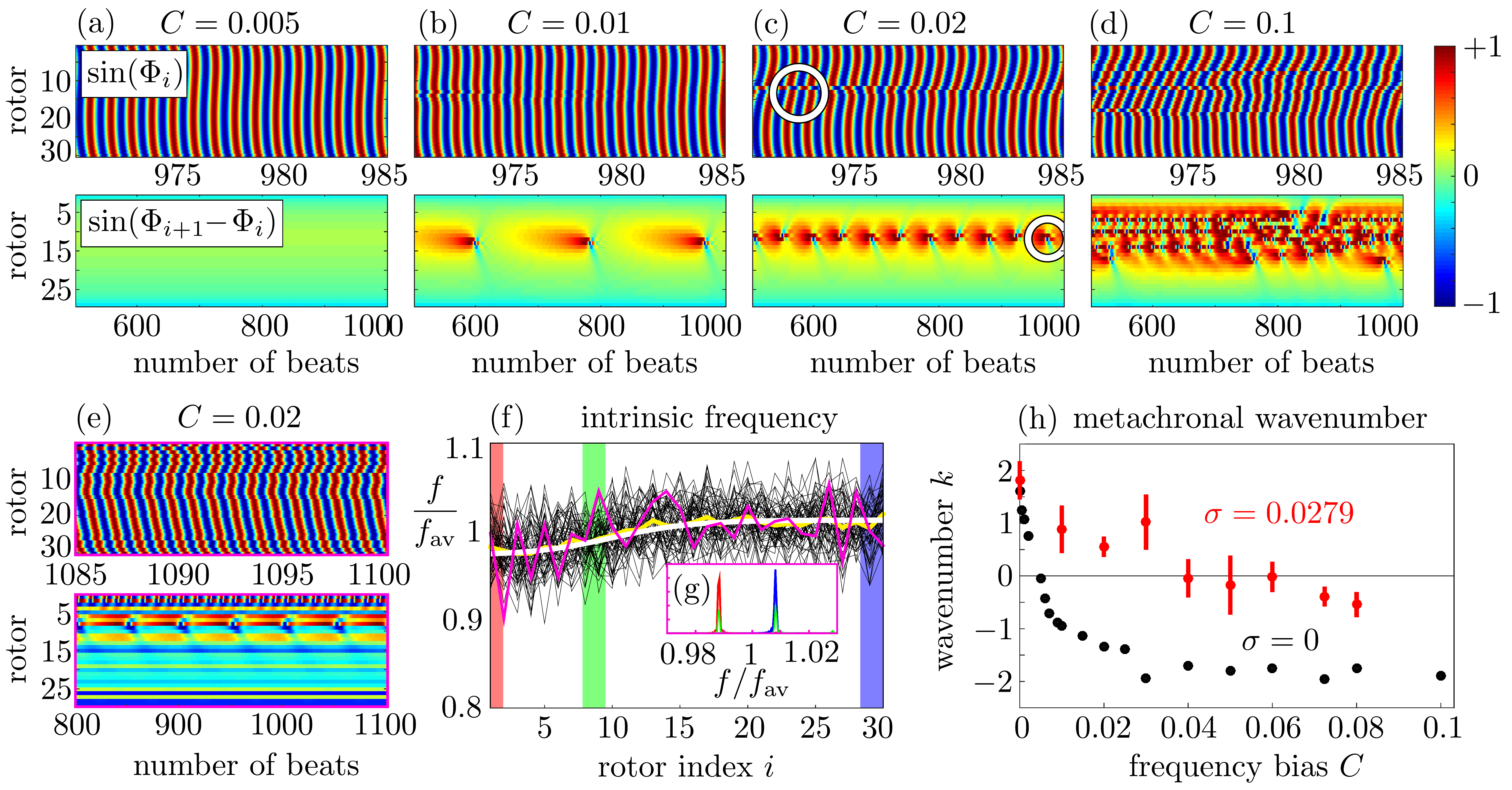}
\caption{Inhomogeneous driving forces and the appearance of defects. Rotor chains possess a {\it Volvox}-inspired intrinsic driving force $f_i^{\text{drive}}/f_0 = 1+C \times(\tanh(0.1684\times i - 1.6813) - 0.3604) + \epsilon_i$ where $\epsilon_i = \epsilon_i(\sigma)$ is a random number chosen from a normal distribution with mean of zero and standard deviation $\sigma$. The frequency bias along the chain is governed by the parameter $C$. (a-d) Kymographs showing the phase, $\sin (\Phi_i)$, and phase difference, $\sin(\Phi_{i+1}-\Phi_i)$, for various values of $C$, with $\sigma=0$ (no polydispersity). An example phase defect is circled in white. (e) Kymographs for $C=0.02$ with $\sigma=0.0279$. (f) Intrinsic frequency distributions (black) for 60 different realisations with $C=0.02$ with $\sigma=0.0279$, along with their average (yellow). The magenta curve corresponds to the frequency profile used in (e) and the inset (g) shows power spectra of $\sin(\Phi_i)$ for oscillators 1 (red), 8-9 (green) and 30 (blue) in this chain. (h) Extracted metachronal wavenumber $k$. Simulations without polydispersity ($\sigma=0$, black, 19 simulations) and with polydispersity ($\sigma=0.0279$, red, 210 simulations) are shown. Other parameters are given by $a/d = 0.01$, $r_0/d = 0.5$, $l/d=2$ and $\Lambda_i$ centred at $\Lambda = 0.1$.}
\label{fig8}
\end{center}
\end{figure*}

{\bf Inhomogeneous driving forces and defects.---}
The experimental results in Fig.~\ref{fig2}(g) demonstrate a spatially-dependent intrinsic 
frequency bias among the flagella in {\it Volvox}. The average shape of this bias is 
captured very well by the functional form $f/f_{\text{av}} = 1+a[\tanh(b(\theta-c))-\tanh(b(d-c))]$, 
chosen so that the average frequency in $\theta \in [0,\pi]$ is independent of the amplitude $a$. 
For a chain of 30 rotors, we utilise the fitted driving force 
$f_i^{\text{drive}} / f_0 = 1+C [\tanh(0.1684\times i - 1.6813) - 0.3604] + \epsilon_i$. The parameter $f_0$ is chosen to centre the distribution of resulting effective spring constants 
$\Lambda_i = \lambda d/f_i^{\text{drive}}$ around $\Lambda=0.1$. The amplitude $C$ is a 
parameter controlling the extent of the frequency bias. 
The value $\epsilon_i = \epsilon_i(\sigma)$ is a random number chosen for each 
rotor from a normal distribution with mean of zero and standard deviation $\sigma$, and 
represents polydispersity in the driving forces along the chain. The appropriate parameters for 
{\it Volvox}, $C_{V}=0.0724$ and $\sigma_V = 0.0279$, were fitted directly from Fig.~\ref{fig2}(g). 
The dynamics of rotor chains are explored with $0 \leq C \leq 0.1$ and $\sigma \in \{0,0.0279\}$, 
the results of which are shown in Fig.~\ref{fig8}.

For sufficiently small frequency biases ($C \leq 0.007$), the system still converges to a 
phase-locked state, and the only effect is to distort slightly the MW (see Fig.~\ref{fig8}(a)). 
This shows that the MW is robust, a result further confirmed by simulations presented in the 
supplementary materials, which explore independently the effects of unbiased polydispersity and 
a linear frequency bias. For $C > 0.007$, the effect of the bias is the emergence of defects 
concentrated around the region of maximal slope in the frequency profile (see Figs.~\ref{fig8}(b-d)). 
These defects are situated in the central/upper part of the chain -- the same region as for the 
experimental results in Fig.~\ref{fig2}(a) -- but exhibit somewhat slower dynamics than the experimental 
defects, lasting approximately 5 to 6 beats, compared to 3 to 4 for the experiments. The defects separate 
two regions of symplectic metachronal coordination, a small one in the upstream portion of the chain 
and a larger one in the downstream half. In particular, it is evident that the bottom third of the 
strip, corresponding to the region of smallest bias, is always coordinated along a symplectic MW. 
This reflects our experimental observations (see Fig. 2(a)), and it is true for $C \leq 0.02$. 
Above that value, this region tends to shrink, going to just 6 oscillators for $C=0.08$. For values 
of $C$ large enough to induce defects but $\leq 0.03$ the power spectrum of individual oscillators 
shows the same qualitative behaviour observed in the experiments: two well defined frequencies for 
the top and bottom groups of phase-locked oscillators, and in the middle a transition region 
characterised by power spectra with a double peak. The relative power carried by the two peaks 
depends on how close the oscillator is to one or the other of the phase-locked groups. The 
transition region is where the defects appear, as is the case for the experiments as well. 
Altogether, these results suggest that $C=0.02$ is an appropriate choice to mimic experimental 
behaviour in the model.

\begin{figure*}[t!]
\begin{center}
\includegraphics[width=0.9\textwidth]{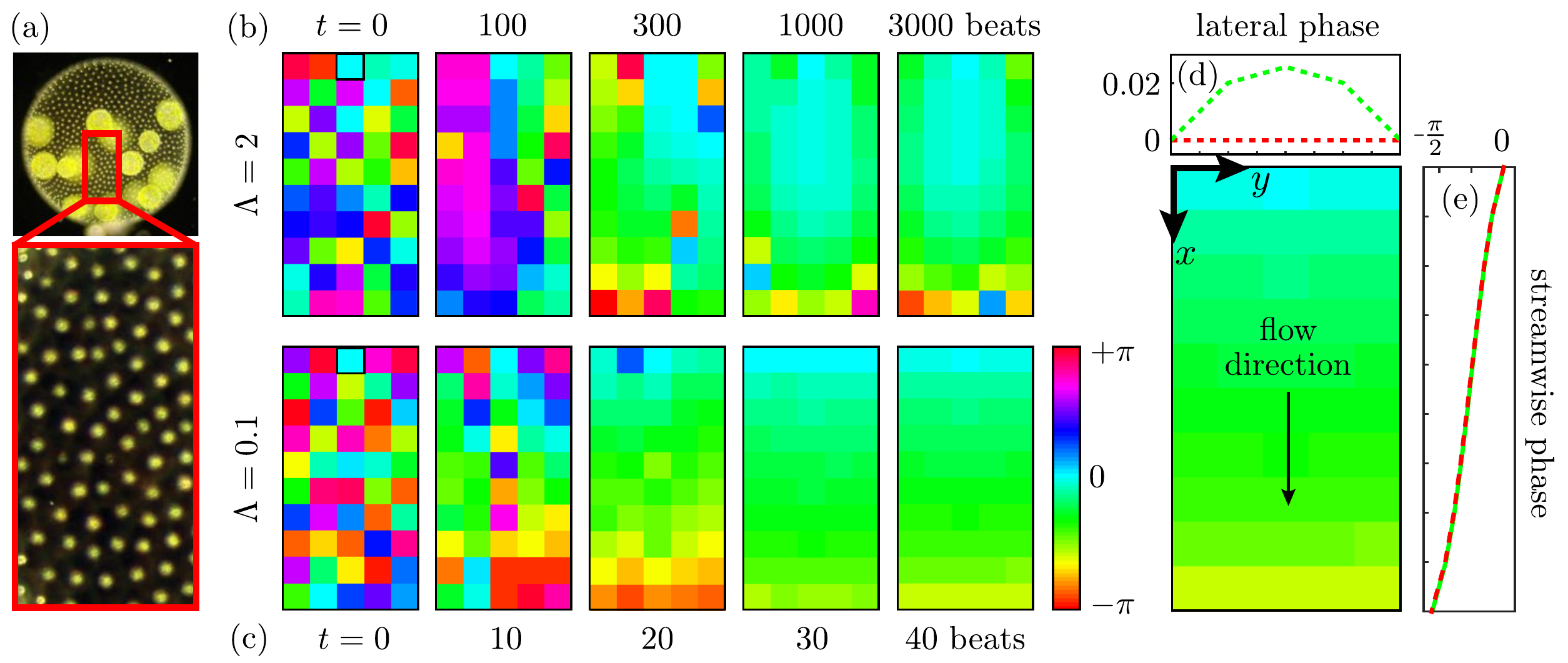}
\caption{Two-dimensional lattice. (a) Photograph of a patch of flagella on the surface of {\it Volvox} (Goldstein). Phase 
evolution for a two-dimensional square grid of 50 rotors with  (b) $\Lambda = \lambda d/f^{\text{drive}} = 2$ and (c) the 
{\it Volvox}-like parameter $\Lambda = 0.1$. In each case, the phase is shown relative to that of the top, central rotor. Other 
parameters are: orbit radius $r_0/d = 0.5$, rotor-rotor spacing $l/d=2$ and sphere size $a/d = 0.01$. Simulations in b-c have been 
conducted with free boundary conditions. (d) Average lateral and (e) streamwise phase profiles for free boundary conditions (green 
dashed) and periodic boundary conditions in the $x$-direction (red dashed).}
\label{fig9}
\end{center}
\end{figure*}

The effect of including polydispersity in the driving forces was also investigated, using the measured value from {\it Volvox} 
($\sigma = \sigma_V = 0.0279$) in the simulations. For various values of $C$, sets of simulations were conducted, each a different 
realisation of this random polydispersity (210 in total, across 9 values of $C$). Figure~\ref{fig8}(f) shows the intrinsic frequency 
profile for 60 different simulations with $C=0.02$ (black) together with their average (yellow), and the underlying bias (white). The 
magenta curve corresponds to the kymographs presented in Fig.~\ref{fig8}(e). Even in the presence of polydispersity, the system displays 
a symplectic MW on average. 

The correlation function used to characterise the mean experimental wave properties is also utilised here to measure the average metachronal wavenumber in the numerical simulations. For each simulation, the wavenumber $k$ was extracted, the results of which are shown in Fig.~\ref{fig8}(h). Without polydispersity ($\sigma=0$), increasing the frequency bias $C$ results in a reversal of the metachronal wave, with $k \sim -2$ for $C \geq 0.03$. However, including random variations in the intrinsic rotor frequencies ($\sigma = 0.0279$) increases the wavenumber significantly. For $C=0.03$, the metachronal wave remains symplectic, with wavenumber $k \sim 1$. Incorporating polydispersity in this fashion increases the frequency of phase defects, which allow relaxation of the rotor chain to the locally-preferred symplectic state. The repeated disruption of this metachronal wave due to polydispersity in the rotor properties allows the wave to sustain clear symplectic coordination on average, even in presence of a frequency bias that would otherwise induce a strong reversal of wave direction.

{\bf Two-dimensional array of rotors.---} Figure~\ref{fig9}(a) shows a patch of {\it Volvox} cells on the 
colony's surface. Although the distribution is not entirely regular, the intercellular spacing is almost uniform. Following 
this observation, we will investigate in this section the dynamics of a two-dimensional array of $10 \times 5$ identical rotors 
($\hat{x}$ and $\hat{y}$ axes respectively), arranged on a square lattice of constant spacing  $l/d=2$. The rotor 
orbits are parallel to $\hat{x}$, with $r_0/d = 0.5$, $a/d = 0.01$, and a radial stiffness $0.1 \leq \Lambda \leq 2$.

Commencing from random initial phases, the system with $\Lambda = 2$ moves over several hundred beats towards a phase profile in which the interior rotors lead (see Fig.~\ref{fig9}(b)). This is analogous to the profile exhibited for the linear chain in Fig.~\ref{fig7}(c), whereby drag reduction of interior rotors leads to a chevron phase profile. Figure~\ref{fig9}(c) shows that for $\Lambda = 0.1$ the system develops a symplectic MW within a few tens of cycles with average phase difference of $\bar{\Delta}_S = -0.19$ between adjacent rotors with the 
same $y$-coordinate. This is very similar to the value found for a linear chain ($\bar{\Delta}_S=-0.17$), indicating that the one-dimensional configuration is able to capture the salient features of the system. Figures~\ref{fig9}(d-e) show the average lateral and streamwise phase profiles respectively (green dashed curves). 

The same behaviour was observed when imposing periodic boundary conditions along $\hat{y}$, mimicking the periodic 
arrangement of  surface-mounted flagella on {\it Volvox}. Starting from random initial conditions, all our 
simulations for $\Lambda=0.1$ converged rapidly to a symplectic MW with rotors along lines of constant $x$ locking in-phase. There was almost no measurable effect on the streamwise metachronal wave profile (red dashed curve in Fig.~\ref{fig9}(e)), though the slight phase variations in the $y$-direction were suppressed. This periodic arrangement can also sustain a discrete set of lateral MW components. Our simulations show that a MW with a total phase accumulation of $\pm 2 \pi$ along the periodic direction ($\hat{y}$) is locally stable, 
and higher windings $\pm 2 \pi n$ should also be possible, although their domain of stability is likely to shrink with 
increasing integer $n$. These states  did not develop in any of our simulations starting with random initial conditions, 
so the probability to realise them over the purely symplectic wave is likely to be very small. 

\section*{Conclusions}

This paper presents a detailed experimental analysis of the global flagellar dynamics on the surface of the 
multicellular green alga {\it Volvox carteri}. 
As a model organism, {\it Volvox} has the fundamental advantage that its flagella are well separated from each 
other and hence much more in the low-coupling limit than for any other organism where metachronal coordination 
has been previously reported (e.g. protozoa, ciliated epithelia). This property allowed us to compare successfully 
flagellar dynamics in {\it Volvox} with a minimal model of hydrodynamically coupled elastic rotors. 
The model develops spontaneously a symplectic MW of a wavenumber that compares well with the one observed experimentally. 
This MW is robust against inhomogeneities in the rotors' intrinsic frequencies, variations in their spatial arrangement, and choice of boundary conditions. These are all essential properties for a plausible model of metachronal coordination. We have shown the existence of MWs which are consistently symplectic on average, but display at the same time characteristic recurrent defects. These might be connected to the observed bias in the distribution of beating frequencies across the colony, which by itself would favour a wave propagating in the opposite direction. We tested this hypothesis within our minimal model, and observed the emergence of recurrent phase defects in the same location and with the same power spectral signature as those observed experimentally.
Altogether, these results show that different properties of the underlying oscillators can combine and give rise to much more complex metachronal wave dynamics than previously assumed.

\section*{Acknowledgements}

We are grateful to D. Page-Croft, C. Hitch, and J. Milton for technical assistance.  

\section*{Funding Statement}

This work was supported in part by the EPSRC (MP), ERC Advanced Investigator Grant 247333, 
and a Senior Investigator Award from the Wellcome Trust.

\newpage

\noindent{\large {\bf Supplementary Material}}
\medskip

The results of various simulations are shown here for chains of rotors above a no-slip wall, with inhomogeneities in their driving forces. We explore the effects of introducing a frequency bias of the same functional form as the measured frequency distribution in {\it Volvox} (see Fig.~2(g)), and also additional polydispersity in the rotor driving forces. This frequency profile in Fig.~2(g) was fitted and for a chain of 30 rotors, the appropriate driving force is given by $f_i^{\text{drive}}/f_0 = 1+C[\tanh(0.1684\times i - 1.6813) - 0.3604]$ with $C_{\textit{V}}=0.0724$ and $1 \leq i \leq 30$ is the rotor index. The parameter $f_0$ is chosen to centre the distribution of resulting effective spring constants $\Lambda_i = \lambda d/f_i^{\text{drive}}$ around $\Lambda=0.1$. Figure~2(g) also shows that the frequency profiles for individual {\it Volvox} colonies are distributed around the mean curve. The standard deviation of this spread was measured to be $\sigma_V=0.0279$ for {\it Volvox}. In the following sections, various driving forces and degrees of polydispersity are considered. Multiple realisations of each parameter set are shown, highlighting the general behaviour of each configuration.

\newpage
\section*{Increasing frequency bias ($0.001 \leq C \leq 0.1$) \\ No polydispersity ($\sigma=0$)}

For small values of $C$, the shape of the MW is perturbed, yet remains phase-locked and symplectic in nature. For $C>0.007$ phase defects begin to emerge among the interior rotors. These defects are periodic for $0.007 < C < 0.02$, but for larger values of $C$, defects emerge at various points in the chain and interact to produce more complex phase dynamics. In every simulation, there exists a region of oscillators at each end which are locally phase-locked. 

\begin{figure}[h!]
\begin{center}
\includegraphics[width=\columnwidth]{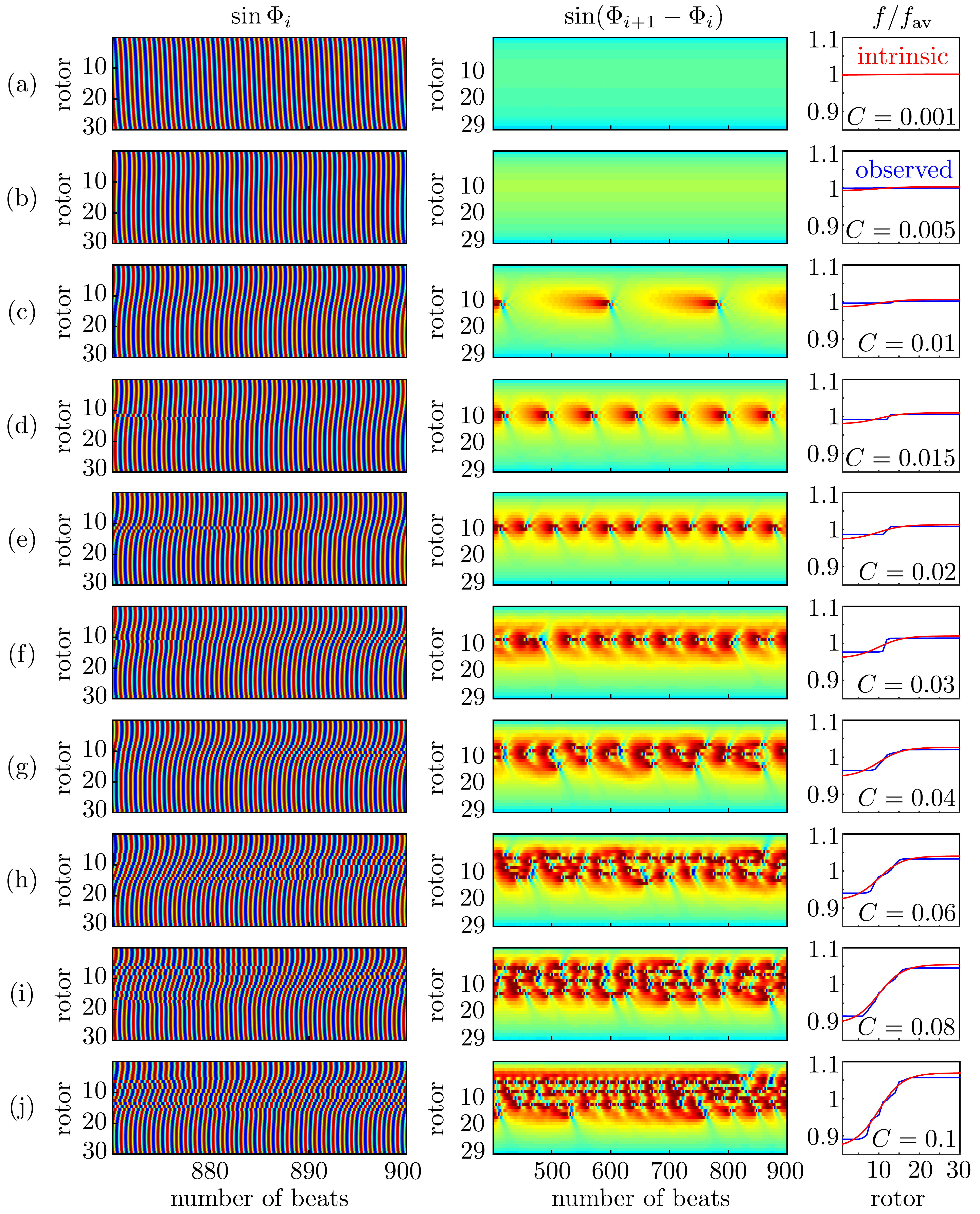}
\caption{Chains of rotors with an intrinsic frequency bias. The driving force along each chain is given by $f_i^{\text{drive}} / f_0 = 1+C[\tanh(0.1684\times i - 1.6813) - 0.3604]$ for various values of $C \in [0.001,0.1]$. The value for {\it Volvox} is given by $C_{V}=0.0724$. Other parameters are given by $\Lambda = 0.1$, $a/d = 0.01$, $r_0/d = 0.5$, $l/d=2$. The phase profile and phase difference from each simulation are shown, as well as the {\it intrinsic} rotor frequency profile (red curves) and average {\it measured} frequency profile from the simulations (blue curves). As the frequency bias is increased, defects in the middle of the chain emerge, becoming more abundant for larger values of $C$.}
\label{S1}
\end{center}
\end{figure}

\newpage
\section*{Fixed frequency bias ($C = 0.02$) \\ Polydispersity ($\sigma=0.0279$)}

\begin{figure}[h!]
\begin{center}
\includegraphics[width=\columnwidth]{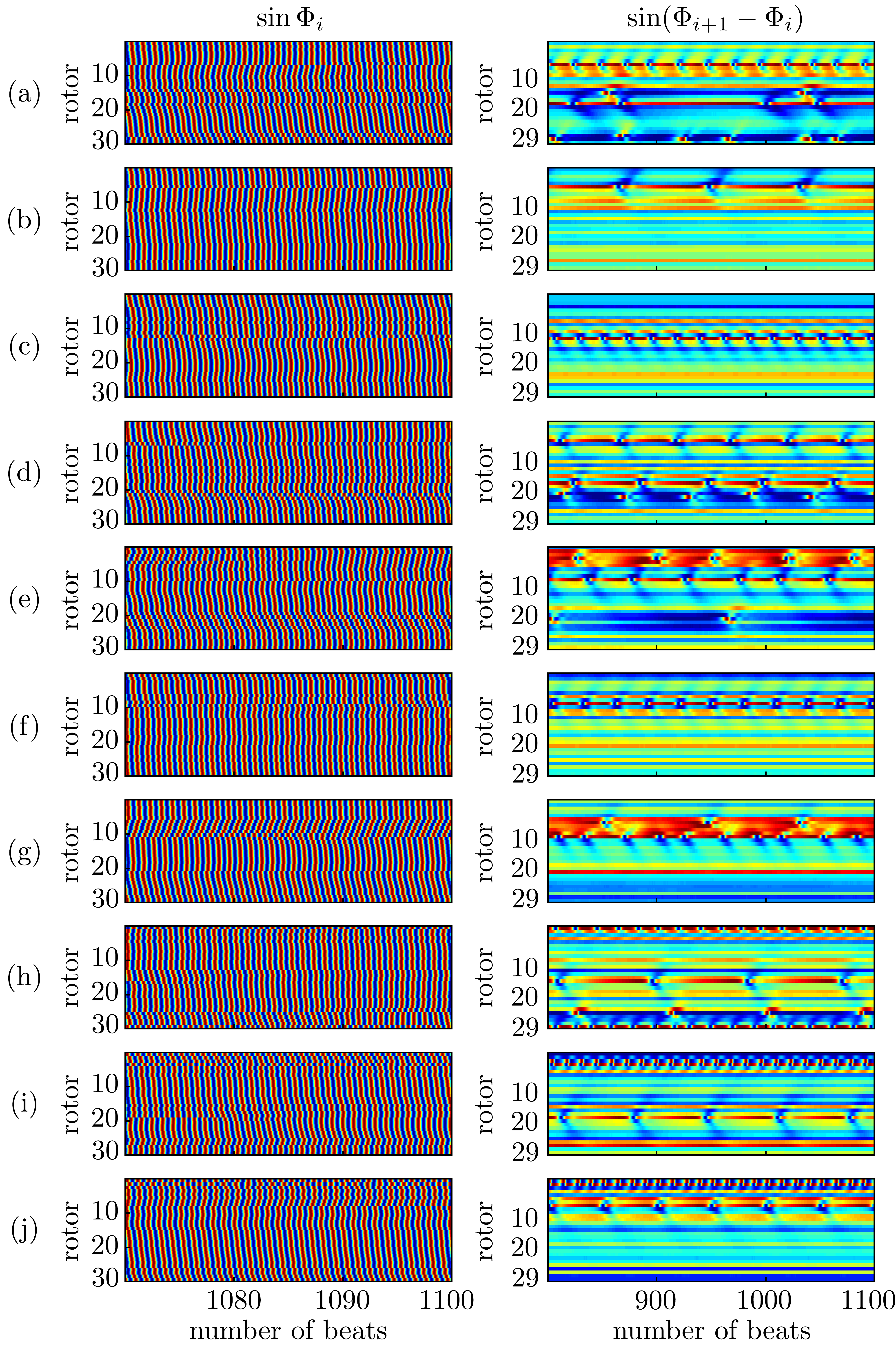}
\caption{Various realisations of polydispersity. The driving force along each chain is given by $f_i^{\text{drive}} / f_0 = 1+0.02[\tanh(0.1684\times i - 1.6813) - 0.3604] + \epsilon_i$ where $\epsilon_i = \epsilon_i(\sigma)$ is a random number chosen from a normal distribution with mean of zero and standard deviation $\sigma=\sigma_V = 0.0279$. Other parameters are given by $\Lambda = 0.1$, $a/d = 0.01$, $r_0/d = 0.5$, $l/d=2$. The phase profile and phase difference in each simulation are shown. The colour scale includes $-1$ (blue) through to $+1$ (red). The results corresponding to 10 of the 60 simulations are shown here, and are representative of the general behaviour observed.}
\label{S2}
\end{center}
\end{figure}

\newpage
\section*{Fixed frequency bias ($C = 0.0724$) \\ Polydispersity ($\sigma=0.0279$)}

\begin{figure}[h!]
\begin{center}
\includegraphics[width=\columnwidth]{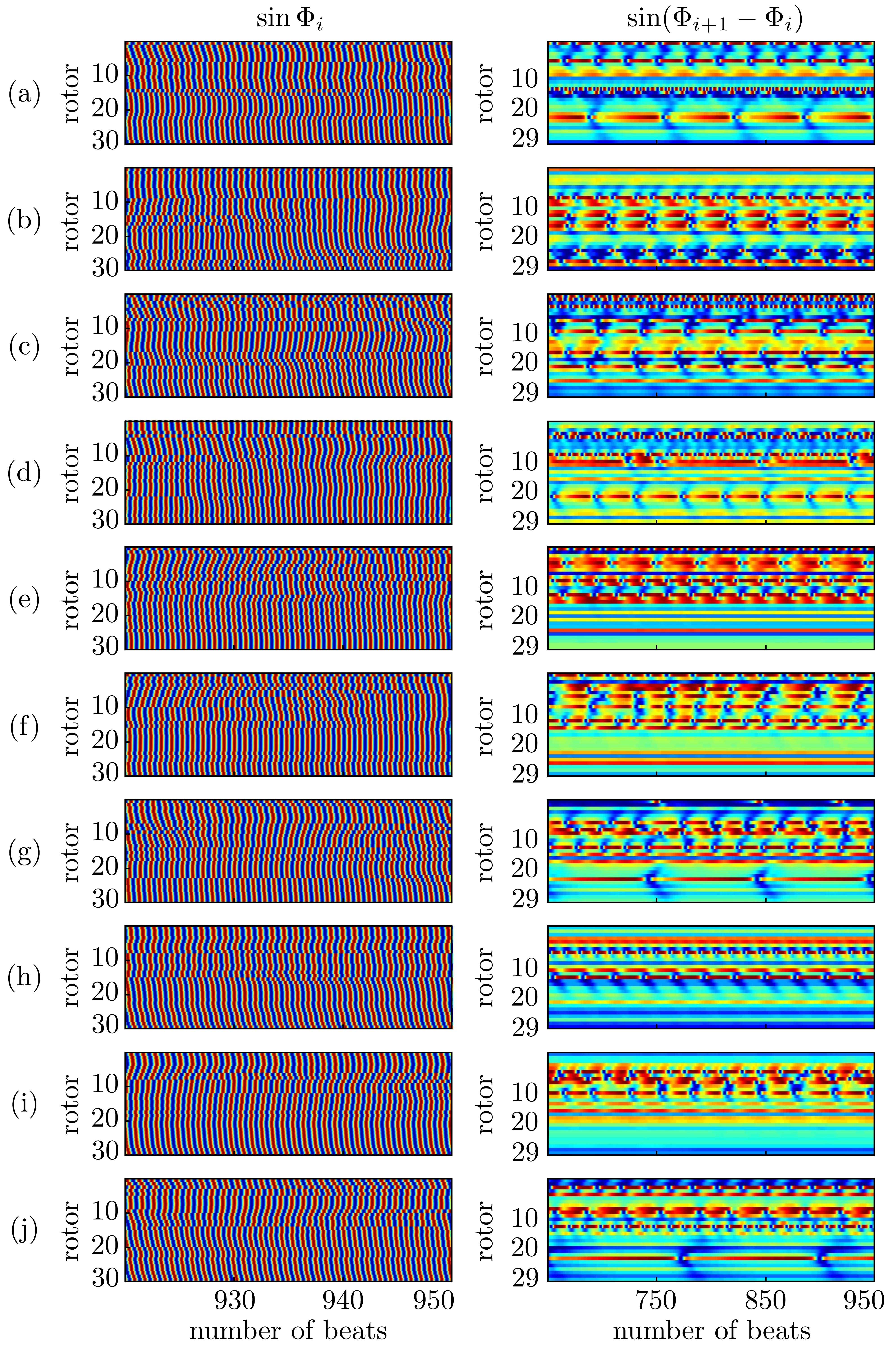}
\caption{Various realisations of polydispersity. The driving force along each chain is given by $f_i^{\text{drive}} / f_0 = 1+C[\tanh(0.1684\times i - 1.6813) - 0.3604] + \epsilon_i$ where $C=C_V=0.0724$ and $\epsilon_i = \epsilon_i(\sigma)$ is a random number chosen from a normal distribution with mean of zero and standard deviation $\sigma=\sigma_V = 0.0279$. Other parameters are given by $\Lambda = 0.1$, $a/d = 0.01$, $r_0/d = 0.5$, $l/d=2$. The phase profile and phase difference in each simulation are shown. The colour scale includes $-1$ (blue) through to $+1$ (red). The results corresponding to 10 of the 60 simulations are shown here, and are representative of the general behaviour observed.}
\label{S3}
\end{center}
\end{figure}

\newpage
\section*{No frequency bias ($C = 0$) \\ Polydispersity ($\sigma=0.0279$)}

\begin{figure}[h!]
\begin{center}
\includegraphics[width=\columnwidth]{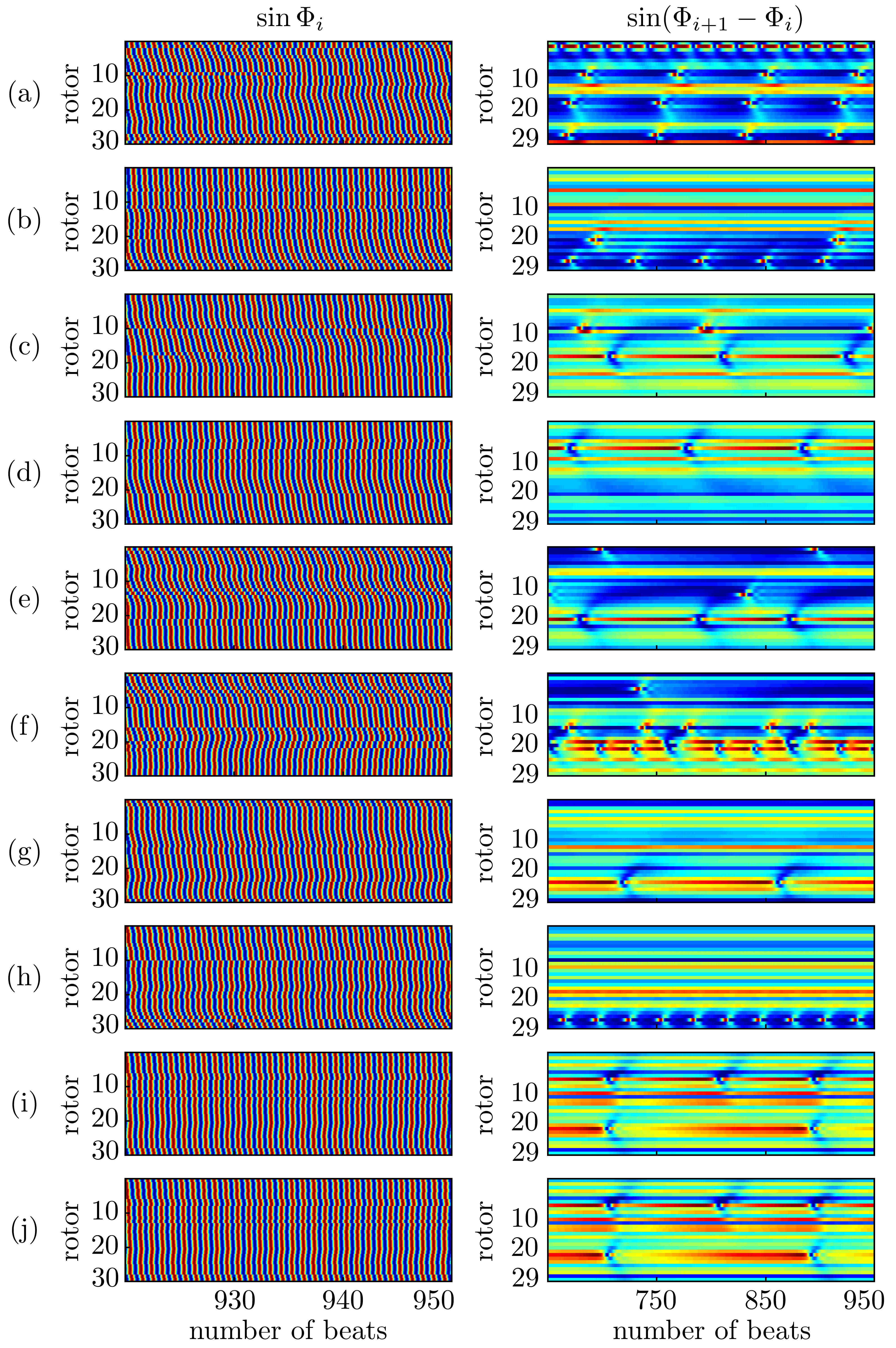}
\caption{Various realisations of polydispersity. The driving force along each chain is given by $f_i^{\text{drive}} / f_0 = 1+ \epsilon_i$ where $\epsilon_i = \epsilon_i(\sigma)$ is a random number chosen from a normal distribution with mean of zero and standard deviation $\sigma =\sigma_V = 0.0279$. Other parameters are given by $\Lambda = 0.1$, $a/d = 0.01$, $r_0/d = 0.5$, $l/d=2$. The phase profile and phase difference in each simulation are shown. The colour scale includes $-1$ (blue) through to $+1$ (red). The results corresponding to 10 of the 30 simulations are shown here, and are representative of the general behaviour observed. Symplectic MWs persist on average, even in the presence of polydispersity.}
\label{S4}
\end{center}
\end{figure}

\newpage
\subsubsection*{Correlation parameters}

In this section, the correlation parameters are extracted for chains of rotors with varying degrees of polydispersity. The driving force for the $i^{\text{th}}$ rotor is taken to be $f_i^{\text{drive}}/f_0=(1 + \epsilon_i)$ where $\epsilon_i = \epsilon_i(\sigma)$, as in Fig.~\ref{S4}, but for various values of $\sigma$. The parameter $f_0$ is chosen to centre the distribution of resulting effective spring constants $\Lambda_i = \lambda d/f_i^{\text{drive}}$ around $\Lambda=0.1$. With these parameters, the threshold detuning for an isolated pair is $|\Omega|_{\text{max}}\simeq 6\%$. Here we want to investigate the extent to which this applies to MWs, and extract the correlation parameters for the emergent metachronal waves.

Altogether, we performed 150 simulations of arrays of 30 rotors, each run for approximately 3000 cycles starting 
with random initial conditions. For each simulation, the limit cycle of individual rotors was independently 
determined, and used to define their phases and beating periods $\{\Phi_i,T_i\}_{i=1}^{30}$. The normalised 
standard deviation of the set of periods $\psi = \text{std}(\{T_i\}_{i=1}^{30}) / T_{\text{av}}$ characterises the 
variation in the set of beating periods for each simulation. Here $T_{\text{av}}$ is the average of $\{T_i\}$.  As $\psi$ is increased, the MW that would develop in a system of identical rotors is perturbed initially 
only slightly ($\psi=0.009$; rotors globally phase-locked), then more heavily ($\psi=0.028$; rotors locally phase-locked), until eventually the MW is realised only on average and any locking is only local ($\psi=0.052$). Kymographs were used to estimate the average MW properties $(T,\tau,L,k)$ for each simulation, as described 
previously for the experiments.

\begin{figure}[t!]
\begin{center}
\includegraphics[width=0.9\columnwidth]{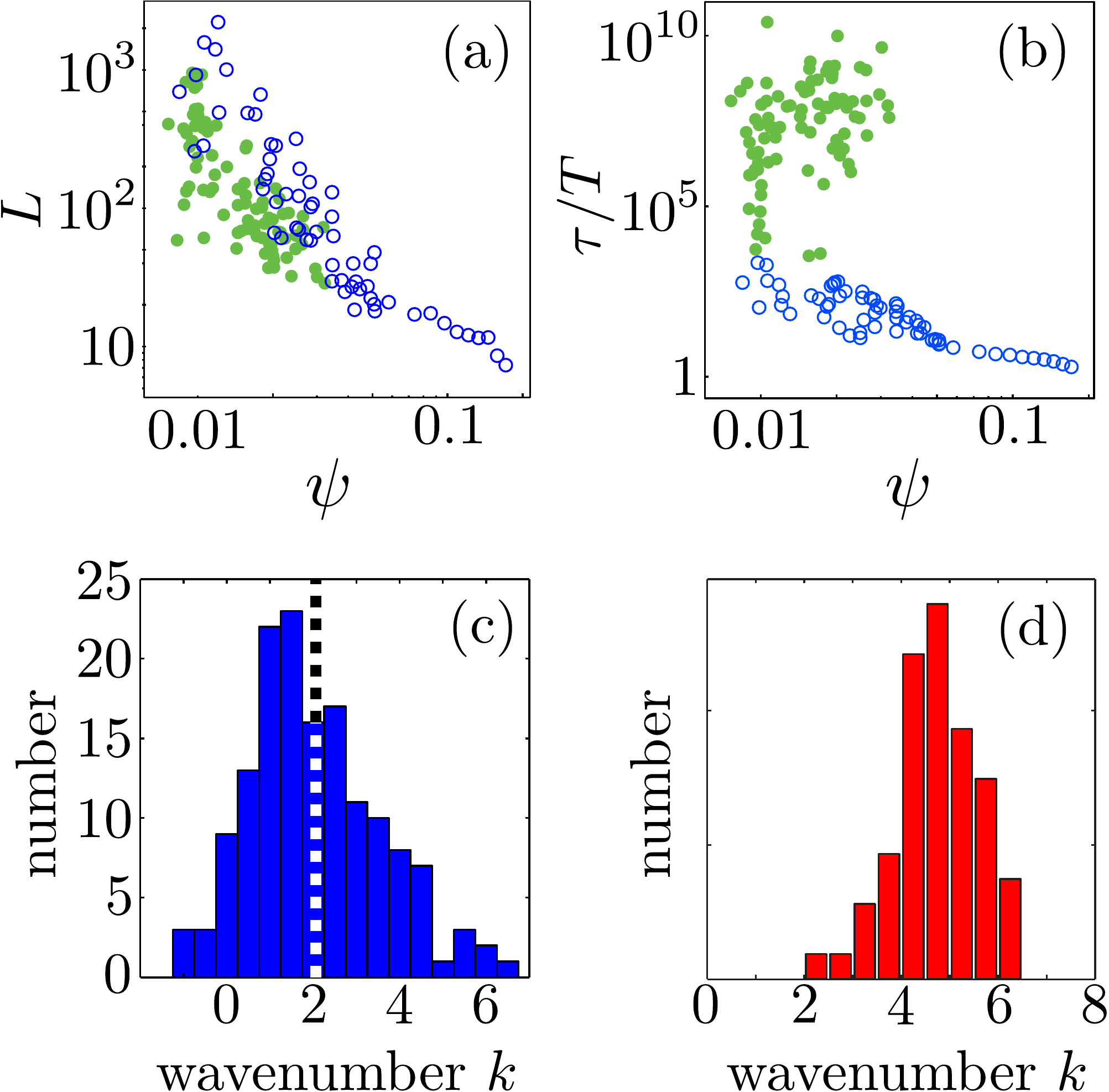}
\caption{(a) Correlation decay length $L$ and (b) time $\tau/T$ scaled by period are shown as functions of the spread in beating periods $\psi$. (c) Histogram of extracted wavenumbers. Values of $\psi$ are the same as the preceding two plots. Results correspond to 150 simulations of $N=30$ spheres with $a/d = 0.01$, $r_0/d = 0.5$, $l/d=2$ and $\Lambda_i$ centred at $\Lambda = 0.1$. (d) Histogram of wavenumbers in {\it Volvox} colonies \cite{Brumley:2012}. }
\label{S5}
\end{center}
\end{figure}

Figures~\ref{S5}(a-b) show that, generally, both the autocorrelation length and time, $L$ and 
$\tau$, decrease significantly as $\psi$ is increased. In these plots, the systems with $\tau$ larger than the duration of 
the simulation are coloured green, while the remaining points are shown in blue. The autocorrelation time 
plot reveals that these two groups cluster around markedly distinct values of $\tau$. Simulations in the first 
group have $\tau/T\sim 10^6-10^7$ and seemingly independent to the value of $\psi$.  They develop a 
steady MW despite the 
finite dispersity, but only up to $\psi\sim3\%$. This value corresponds to the threshold detuning we encountered for 
two rotors, because for a pair $\psi=\Omega/2$. In the second group, $\tau$ decreases with 
$\psi$. Figure~\ref{S5}(c) displays the spread of wavenumbers $k$. The mean wavenumber ($k=2.1$, dashed line) compares well with the value for identical rotors. With a finite dispersity in the driving forces, 
the wavenumbers spread in a manner similar to the distribution observed experimentally for 
{\it Volvox} (Fig.~\ref{S5}(d)).  A few systems at large values of $\psi$ displayed 
negative wavenumbers ($\sim6\%$ of the simulations), but overall the simulations show that the present 
model robustly exhibits symplectic MWs, even with variations in rotor properties.

\newpage
\section*{Linear frequency bias \\ No polydispersity ($\sigma=0$)}

\begin{figure}[h!]
\begin{center}
\includegraphics[width=0.9\columnwidth]{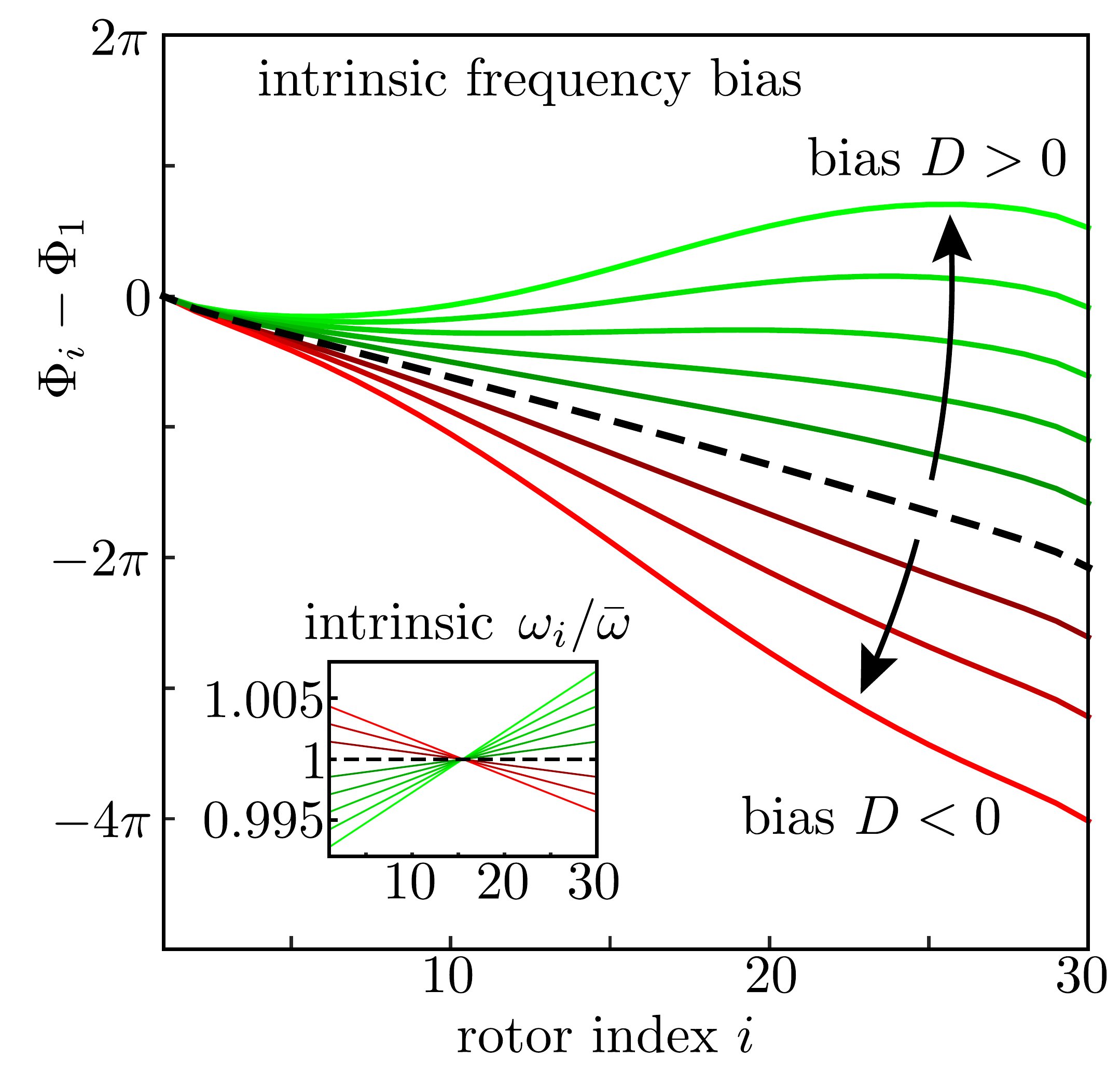}
\caption{Steady state metachronal wave for chains of rotors with an intrinsic frequency bias. Inset shows the intrinsic frequency of rotors along each chain. The functional form of the driving force is linear, and given by $f_i^{\text{drive}} / f_0 = 1 + D\big[ i - (N+1)/2 \big]$ for various values of $D \in [-0.0003,+0.0005]$. Other parameters are given by $\Lambda = 0.1$, $a/d = 0.01$, $r_0/d = 0.5$, $l/d=2$.}
\label{S6}
\end{center}
\end{figure}

To this point, the {\it Volvox}-inspired intrinsic frequency profile has been used, with the parameter $C$ controlling the extent of this bias along the chain. Chains of rotors with a {\it linear} bias in their intrinsic frequencies are studied here. The rotors are actuated with a driving force of the form $f_i^{\text{drive}}/f_0 = 1 + D\big[ i - (N+1)/2 \big]$ for various values of $D \in [-0.0003,+0.0005]$, so that their intrinsic beat frequency varies monotonically along the chain. Figure~\ref{S6} shows the steady state phase profiles exhibited for a number of values of $D$. This frequency bias can either enhance ($D<0$) or reduce ($D>0$) the slope of the metachronal wave, while still permitting convergence to a steady state. The direction of the metachronal wave is robust, even when the constituent rotors possess a mild opposing frequency bias ($\sim 1$\% difference between the end rotors).

There are significant qualitative differences between this system, and the one presented in Fig.~\ref{S1}. For the linear profile here, a very small overall frequency difference between the end rotors ($D=+0.0005$) results in suppression of the symplectic metachronal wave. Effective hydrodynamic coupling for each rotor does not extend to more than a few neighbours away, and so the mild linear frequency profile does not provide ``weak points'' in the coupling. Conversely, the intrinsic frequency profiles in Fig.~\ref{S1} vary nonlinearly. Defects emerge in the middle of the rotor chain where the neighbouring frequency difference is largest, allowing relaxation to the symplectic metachronal wave.

\end{document}